\documentclass[aps, prx, amsmath, superscriptaddress,preprintnumbers,twocolumn]{revtex4-2}
\usepackage{graphicx}
\usepackage{dcolumn}
\usepackage{bm}
\usepackage{amsmath}
\usepackage{subfigure}
\usepackage{amsfonts}
\usepackage[svgnames]{xcolor}
\usepackage{appendix}
\usepackage{multirow}
\usepackage{booktabs}
\usepackage{tabularx}
\usepackage[colorlinks,
            linkcolor=blue,
			filecolor=blue,
			urlcolor= blue,
			citecolor=blue,
            ]{hyperref}
\usepackage{amsthm}
\usepackage{setspace}
\usepackage[T1]{fontenc}
\usepackage{extarrows}
\usepackage{lineno}

\usepackage[percent]{overpic}
\usepackage{dsfont}
\usepackage{floatrow}
\usepackage{physics}

\newcommand{\kett}[1]{\left.\ket{#1}\right\rangle}

\newcommand{\brakett}[2]{\left\langle\braket{#1}{#2}\right\rangle}
\newcommand{\ee}{{e}}
\newcommand{\ii}{{i}}

\newcommand{\phii}[0]{\hat{\Phi}}
\newcommand{\mm}[0]{\hat{\mathcal{M}}}
\newcommand{\M}[0]{{\mathcal{M}}}

\newcommand{\vk}{{\bar v_k}}
\newcommand{\vv}{{\bar v}}
\begin{document}
	
\title{Channel-based framework for phase esimation of multiple eigenvalues}

\author{Yuan-De Jin}
\affiliation{State Key Laboratory of Superlattices and Microstructures, Institute of Semiconductors, Chinese Academy of Sciences, Beijing 100083, China}
\affiliation{College of Materials Science and Opto-Electronic Technology, University of Chinese Academy of Sciences, Beijing 100049, China}
\affiliation{Department of Applied Physics, University of Science and Technology Beijing, Beijing 100083, China}

\author{Shi-Yu Zhang}
\affiliation{Department of Physics, University of Michigan, Ann Arbor, Michigan 48109, USA}


\author{Wen-Long Ma}
\email{wenlongma@semi.ac.cn}
\affiliation{State Key Laboratory of Superlattices and Microstructures, Institute of Semiconductors, Chinese Academy of Sciences, Beijing 100083, China}
\affiliation{College of Materials Science and Opto-Electronic Technology, University of Chinese Academy of Sciences, Beijing 100049, China}
\date{\today }
\begin{abstract}
	Quantum phase estimation (QPE) of the eigenvalues of a unitary operator on a target quantum system is a crucial subroutine in various quantum algorithms. Conventional QPE is often expensive to implement as it requires a large number of ancilla qubits and the ability to perform quantum Fourier transform. Recent developments in iterative QPE reduce the implementation cost by repetitive uses of a single ancilla and classical post-processing. However, both conventional and iterative schemes often require preparation of the target system in an eigenstate of the unitary operator, while it remains ambiguous to achieve QPE of multiple eigenvalues with no need of initial state preparation. Here we clarify this issue by developing a theoretical framework based on sequential quantum channels for iterative QPE. We find that QPE of multiple eigenvalues can be efficiently realized for arbitrary initial target system state by actively utilizing the measurement backaction of iterative QPE on the target system with a long coherence time. Specifically, we investigate two iterative QPE schemes based on sequential Ramsey interferometry measurements (RIMs) of an ancilla qubit: (a) the repetitive scheme, which conducts repetitive RIMs to achieve the standard quantum limit in estimating the eigenvalues; (b) the adaptive scheme, which adjusts the parameters of each RIM based on prior measurement outcomes to attain the Heisenberg limit. In both schemes, sequential ancilla measurements generate sequential quantum channels on the target system, gradually steering it to the eigenstates of the estimated unitary operator, while the measurement statistics of the ancilla can reveal the embedded information about its eigenvalues with proper post-processing. We demonstrate the analysis by simulating a central spin model, and evaluate the performance and noise resilience of both schemes.
\end{abstract}

\maketitle

\section{Introduction}
Quantum phase estimation (QPE), designed to extract the eigenvalues of a unitary matrix on a target quantum system, is an important module in quantum algorithms offering exponential speedup, such as Shor's algorithm for integer factorization \cite{Shor1994} and Harrow-Hassidim-Lloyd (HHL) algorithm for solving linear systems of equations \cite{Harrow2009}. The conventional QPE scheme requires complex controlled unitary operations and inverse quantum Fourier transform on quite a few ancilla qubits \cite{Nielsen2010, Cleve1998}. Equivalently, the complex controlled unitary operations can be simplified by preparing a Greenberger-Horne-Zeilinger (GHZ) state of a larger number of ancilla qubits. However, the hardware overhead and complex control make such schemes impractical for near-term small-scale experiments.

Another QPE scheme is to perform repeated measurement on a single ancilla qubit with different time durations to estimate different digits of the eigenvalues, known as iterative QPE algorithm \cite{Kitaev1995,Parker2000, Dobsicek2007,OLoan2010,Smith2022b,Ahnefeld2022}. Some recent works have proposed low-depth quantum circuits for estimating multiple eigenvalues with standard quantum limit (SQL) \cite{OBrien2019, Somma2019}. Other efforts aim at improving the precision of the estimation from the SQL to the Heisenberg limit (HL) with adaptive measurements \cite{Higgins2007, Dutkiewicz2022, Giedke2006, Knill2007,Dobsicek2007}, in which subsequent measurements are modified by previous results and Bayesian estimation \cite{Higgins2009, Bonato2016,Wiebe2016}. Repetitive measurements can also be combined with adaptive or Bayesian control, which can result in more accurate estimation by eliminating the estimation degeneration \cite{Smith2022b,Smith2024}. In particular, a simple quantum circuit with one ancilla qubit and sophisticated classical post-processing procedures can realize HL-limited QPE, useful for ground state energy estimation on early fault-tolerant quantum computers \cite{Dong2022,Lin2022a,Ding2023,Ding2023a}. However, previous conventional and iterative QPE schemes often assume that one can (approximately) prepare eigenstates of a target system, while it remains largely unexplored whether there exist alternative systematic approaches for simultaneous estimation of multiple eigenvalues without initial state preparation of the target system. 

\begin{figure*}[htbp]
\includegraphics[width=18.5cm]{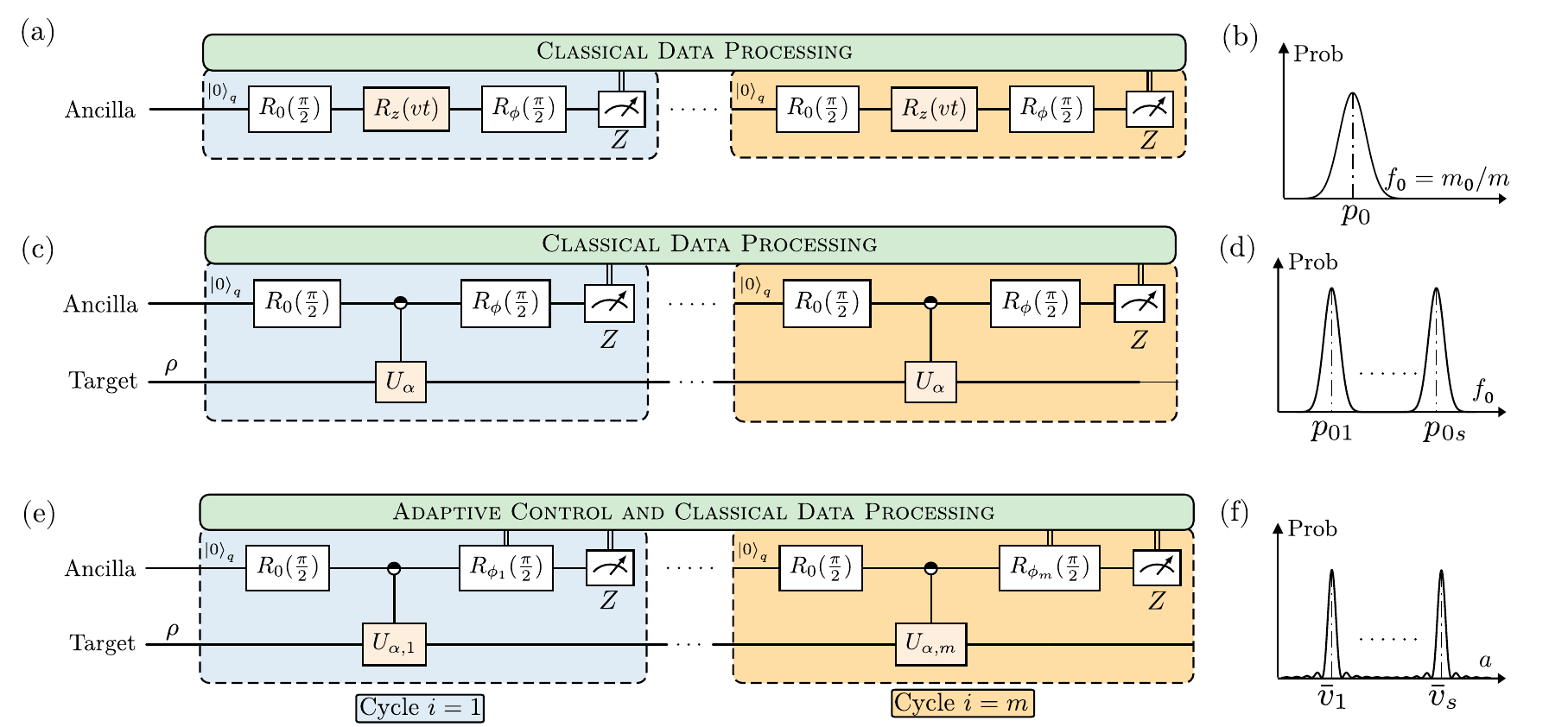}
\caption{ Quantum circuits and measurement statistics of iterative QPE with sequential RIMs. (a) Quantum circuit of repetitive RIMs for estimating a quantity $v$ with an ancilla. Each measurement has two outcomes $\{0, 1\}$. The probability to obtain outcome 0 is $p_0=[1-\cos(2v\tau+\phi)]/2$, and the number of outcome 0 in $m$ measurements is denoted as $m_0$, which is collected by classical data processing. (b) The probability of the frequency $f_0=m_0/m$ of outcome 0 for classical phase estimation, which obeys a binomial distribution around $p_0$. (c) Quantum circuit for repetitive iterative QPE with sequential RIMs, where $U_{\alpha}=e^{-i(-1)^{\alpha}V\tau}$ is a unitary operator of the target system conditioned on the ancilla state $\ket{\alpha}_q$ ($\alpha=0,1$). (d) The probability distribution of $f_0$ for repetitive QPE, which can display multiple binomial distributions around $\{p_{01}, \cdots, p_{0s}\}$ with $p_{0k}=[1-\cos(2v_k\tau+\phi)]/2$. (e) Quantum circuit of adaptive iterative QPE. The rotation angle is initialized as $\phi_1=\pi$, and then undated to $\phi_{i+1}=\pi-2\pi0.0a_{i}\dots a_2 a_1$ by adaptive control, which depends on previous and current outcomes with $a_i$ being the outcome of the $i$th measurement. The duration of each RIM cycle also varies with $\tau_i=2^{m-i}\pi/\|V\|$ and $U_{\alpha,i}=e^{-i(-1)^{\alpha}V\tau_i}$. (f) The probability distribution of $a=0.a_m\dots a_2 a_1$ for adaptive QPE, described by multiple Fej\'{e}r kernels around $\{\vv_1,\cdots,\vv_s\}$ with $\vv_k=v_k/||V||$.   }
\label{fig1}
\end{figure*}

In this paper, we present a channel-based framework that rigorously captures the measurement backaction and statistics in iterative QPE, and provide a unified approach for efficient and robust multi-eigenvalue estimation. For iterative QPE based on sequential Ramsey interferometry measurements (RIMs) of an ancilla qubit, we find that QPE of multiple eigenvalues can be efficiently realized without initial state preparation of the target system when actively utilizing the backaction of sequential ancilla measurements on the target system with a long coherence time. We investigate two particular QPE schemes: (a) the repetitive scheme, including a sequence of repetitive RIMs to reach the SQL [see Fig. \ref{fig1}(c)]; (b) the adaptive scheme, including a sequence of adaptive RIMs to reach the HL [see Fig. \ref{fig1}(e)]. For both schemes, we utilize the channel-based framework to clarify the measurement backaction and statistics of sequential RIMs for arbitrary initial state of the target system.

 With sequential RIMs of the ancilla qubit, the target system is gradually steered to the fixed points of a quantum channel induced by the RIM sequence, corresponding to different eigenstates of the unitary operator to be estimated. The gradual state change of the target system influences the ancilla measurement statistics, which contains information about the eigenvalues of the estimated unitary operator.
 Such embedded information can be revealed by appropriate post-processing of the measurement outcomes $\{a_1,\cdots,a_m\}$ with $a_i\in\{0,1\}$ being a binary number, that is, we can choose a stochastic variable $\xi(a_1,\cdots,a_m)$ as a function of all measurement outcomes and infer the eigenvalues by investigating the distribution of $\xi$. For the repetitive scheme, we choose the variable as the average over all binary measurement outcomes, $\xi=\sum_{i=1}^m (1-a_i)/m$, and find that $\xi$ is distributed according to a summation of multiple binomial distributions [Eq. (\ref{Fq})], with each distribution concentrating around a value determined by an eigenvalue of the estimated unitary operator. While for the adaptive scheme, we choose the quantity with the binary expansion $\xi=a_m\cdots a_1/2^m$, whose distribution can be analogously described by a summation of multiple Fej\'{e}r kernels, with each distribution concentrating around a single eigenvalue but with much narrower width [Eq. (\ref{Eq:Prob.Adap})]. The accuracy of QPE can reach the SQL for the repetitive scheme and HL for the adaptive scheme. 

For a noisy target system, both the repetitive and adaptive schemes can estimate the eigenvalues up to some corrections. The estimation in the noisy system for the repetitive scheme depends on the phenomenon of metastability in sequential quantum channels \cite{Jin2024}, such that one can still approximately estimate multiple eigenvalues when the target system evolves into metastable states of the quantum channel. However, the number of measurement repetitions is limited to a relatively narrow interval that is highly sensitive to the noise intensity. In comparison, the adaptive scheme is more immune to noise in that the estimation error saturates as the noise intensity increases. Numerical simulations show that the stable value of this error for the adaptive scheme at least one order of magnitude lower than the minimum value of the repetitive scheme.

The paper is organized as follows. In Sec. \ref{Sec:Pre}, We first introduce the basics of RIM protocols, and then describe the framework based on sequential quantum channels for both repetitive and adaptive iterative QPE schemes. We thoroughly investigate the repetitive QPE scheme in Sec. \ref{Sec:Repetitive} and the adaptive scheme in Sec. \ref{Sec:adaptive}, revealing the measurement backaction and statistics of both schemes. Finally, we compare the performance and noise-resilience of both QPE schemes for a specific spin-star model in Sec. \ref{Sec:Perf}.

\section{Preliminaries and main results}\label{Sec:Pre}
\subsection{QPE of a classical quantity}
We start by introducting the canonical RIM protocol for a single qubit in quantum sensing \cite{Ramsey1950a, Lee2002a, Taylor2008, Liu2019}, which can be regarded as the simplest iterative QPE to sense a classical quantity (or a single eigenvalue) [Fig. \ref{fig1}{\color{blue}(a)}].
Consider an ancilla qubit directly coupled to a classical physical quantity $v$ with the Hamiltonian
\begin{equation}\label{Hc}
    H=v\sigma_{q}^z=v(|0\rangle_q\langle0|-|1\rangle_q\langle 1|),
\end{equation}
where $\sigma_q^i$ is the Pauli-$i$ operator of the ancilla qubit ($i=x,y,z$). Hereafter we denote the ancilla rotation along the axis in equatorial plane as $R_{\phi}(\theta)=e^{-i(\cos\phi \sigma_q^x+\sin \phi \sigma_q^y)\theta/2}$ with $\phi$ denoting the rotation axis and $\theta$ being the rotation angle.

In a single RIM sequence, the qubit initialized to state $|0\rangle_q$, first undergoes a rotation $R_{0}(\pi/2)$, then evolves with the Hamiltonian in \eqref{Hc} for time $t$, undergoes another rotation $R_{\phi}(\pi/2)$ with the resultant state
\begin{align}\label{wave}
    &R_{\phi}(\pi/2) R_z(v\tau) R_{0}(\pi/2)|0\rangle_q \nonumber \\
    =&\frac{1}{2}\left[ {\begin{array}{*{20}{c}}
{{e^{-iv\tau}-e^{i(\phi+v\tau)} }}\\
{-ie^{i\phi}\left({e^{-iv\tau}+e^{i(\phi+v\tau)} }\right)}
\end{array}} \right]_q,
\end{align}
where $R_z(v\tau)=e^{-iv\tau\sigma_q^z }$. Finally a projective measurement (PM) is performed on the ancilla with the readout basis $\{|0\rangle, |1\rangle\}$. The probability distribution for the outcomes $\{0, 1\}$ is $F'=(p_0,p_1)$ with $p_0=[1-\cos(2v\tau+\phi)]/2$ and $p_1=1-p_0$.

The physical quantity $v$ influences the probability distribution $F'$ of the measurement outcomes in a single RIM. If the measurement outcomes of different RIMs are independent and identically distributed (i.i.d.), by repeating the RIMs and averaging over the measurement results, the quantity $v$ can be accurately estimated within the SQL.
Specifically, for $m$ repetitive RIMs, denote the frequencies of measurement outcomes as $F=(f_0,f_1)$, then $f_0$ obeys a binomial distribution \cite{Bengtsson2006}
\begin{equation}\label{Fc}
    p(f_0)=\frac{m!}{(mf_0)!(mf_1)!}p_0^{mf_0}p_1^{mf_1}\approx e^{-mS(F\|F')},
\end{equation}
where $S(F\|F')=\sum_{i=0,1} f_i\ln (f_i/p_{i})$ is the relative entropy between $F$ and $F'$. For large $m$, the above probability distribution is concentrated around $F'$ as a Gaussian distribution since $S(F\|F')\approx\sum_{i=0,1} (f_i-p_i)^2/(2p_i)=(f_0-p_0)^2/(2p_0p_1)$ [Fig. \ref{fig1}{\color{blue}(b)}]. Therefore, $F'$ can be approximated by the most probable $F$ with the standard deviation $\sqrt{p_0p_1/m}$, and $v$ can be inferred from $F$ with a known free evolution time $t$.



\subsection{QPE of multiple eigenvalues}
For iterative QPE of a unitary operator, we assume that an ancilla qubit is coupled to a $d$-dimensional target quantum system with the Hamiltonian
\begin{equation}\label{Hq}
    H=\sigma_q^z\otimes V=\sum_{k=1}^{s}v_k\sigma_q^z\otimes P_k,
\end{equation}
where $V$ is a Hermitian operator on the target system, $\{v_k\}_{k=1}^s$ is the set of distinct eigenvalues of $V$ ($s\leq d$), and $\{P_k\}_{k=1}^s$ is a set of projection operators satisfying $\sum_{k=1}^sP_k=\mathbb{I}$ with $\mathbb{I}$ being the identity operator on the target system. Without loss of generality, we assume that $V$ is a positive operator with $v_k\geq0$ for any $k$ (see Appendix \ref{App:nega} for discussions about a general Hermitian $V$). The operator norm of $V$, defined as $\|V\|={\rm max}\{\sqrt{\langle\psi|V^{\dagger}V|\psi\rangle}/\sqrt{\langle \psi|\psi\rangle}: |\psi\rangle\neq 0\}$, is equal to the maximum eigenvalue in the set $\{v_k\}_{k=1}^s$.

The objective of QPE is to estimate the eigenvalues $\{e^{iv_k\tau}\}_{k=1}^s$ of the unitary operator $e^{iV\tau}$ for a given $\tau$ or the eigenvalues $\{v_k\}_{k=1}^s$ of $V$, which contains critical information about the target system. Such a QPE problem occurs naturally in quantum sensing \cite{degen2017a}, such as using a nitrogen-vacancy (NV) electron spin to detect some nuclear spins in a spin bath \cite{Cappellaro2012, Zhao2011, Zhao2012,Kolkowitz2012,Taminiau2012,Shi2014,Abobeih2019}.

If the target system is prepared in an eigenstate with eigenvalue $v_k$, the Hamiltonian [Eq. (\ref{Hq})] can be replaced by [Eq. (\ref{Hc})], and we can still estimate $v_k$ by repetitive RIMs, with the same steps as those in sensing a classical quantity.
For the target system in an arbitrary state $\rho$, a single RIM on the ancilla induces a quantum channel on the target system \cite{Wolf2010, Caruso2014, Watrous2018}, which can be represented in the Stinespring representation as \cite{Stinespring1955}
\begin{equation}\label{Stine}
    \Phi(\rho)={\rm Tr}_q[U(\tau)(\rho_q\otimes \rho)U^{\dagger}(\tau)],
\end{equation}
where $U(t)=e^{-i H\tau}=\sum_{\alpha=0,1}|\alpha\rangle_q\langle\alpha|\otimes U_{\alpha}$ with $U_{\alpha}=\sum_{k=1}^{s}e^{-i(-1)^{\alpha}v_k\tau}P_k$, $\rho_q=|\psi\rangle_q\langle \psi|$ with $|\psi\rangle_q=R_{0}(\pi/2)|0\rangle_q$, and ${\rm Tr}_q[\cdot]$ denotes the partial trace over the ancilla.
This quantum channel can also be transformed to the Kraus representation as \cite{Kraus1983}
\begin{equation}\label{Stine}
    \Phi(\rho)=\sum_{\alpha=0,1} \mathcal{M}_{\alpha}(\rho)=\sum_{\alpha=0,1}M_{\alpha}\rho M_{\alpha}^{\dagger},
\end{equation}
with $\mathcal{M_{\alpha}}(\cdot)=M_{\alpha} (\cdot) M_{\alpha}^{\dagger}$ is a superoperator with the Kraus operator $M_{\alpha}=[U_0-(-1)^{\alpha}e^{i\phi}U_1]/2$. Note that $\Phi$ is independent of the second rotation $R_{\phi}(\pi/2)$, while $M_0$ and $M_1$ depend on the phase difference $\phi$ between the rotation axes of $R_{0}(\pi/2)$ and $R_{\phi}(\pi/2)$. The probability to obtain result 0 is $p_0={\rm Tr}(M_0\rho M_0^{\dagger})=[1-\sum_{k=1}^s{\rm Tr}(P_k\rho)\cos(2v_k\tau+ \phi)]/2$.
If the probability distributions of each RIM are assumed to be i.i.d, we can still repeat the RIMs, estimate $\{v_k\}_{k=1}^s$ by recording $p_0(\tau)$ as a function of $\tau$ and then perform Fourier analysis to find the frequency components. This is the method for spectral analysis of a unitary in deterministic quantum computation with one quantum bit (DQC1) \cite{Knill1998,Datta2005,Datta2008,Cable2016}.
However, this method neglects the backaction of ancilla measurements on the target system state in sequential RIMs.

Actually the target system undergoes different state jumps induced by the Kraus operators with different measurement outcomes of the ancilla in a single RIM, which will influence the statistics of the next RIM. So the measurement statistics of sequential RIMs can show non-i.i.d. features, especially for a small target system as demonstrated in \cite{Ma2023,Jin2024,Qiu2024}. The main finding in this paper is that such non-i.i.d. features of measurement statistics can be utilized to efficiently estimate the eigenvalues $\{v_k\}_{k=1}^s$. Specifically we investigate two iterative QPE schemes, including the repetitive scheme and the adaptive scheme.

\subsection{Iterative QPE: repetitive and adaptive schemes}
For iterative QPE schemes, sequential RIMs induce sequential quantum channels on the target system, which can be decomposed as a summation of all stochastic trajectories and the asymptotic behavior is projecting the state to eigenspaces (see Appendix \ref{App:Sequential channel}),
\begin{align}
  \Phi_m\cdots\Phi_2\Phi_1(\rho)&=\sum_{a_1,\cdots,a_m}\M_{a_m}\cdots \M_{a_2}\M_{a_1}(\rho), \nonumber \\
  &\approx\sum_{k=1}^s P_k\rho P_k,
\end{align}
where $\Phi_i=\sum_{a_i}\mathcal{M}_{a_i}$ is the quantum channel for the $i$th RIM. Note that the Kraus operators of different RIMs can be different for the adaptive scheme, to simplify the notation, we only use the subscript $a_i$ in $\mathcal{M}_{a_i}$ to denote such a difference. For the quantum trajectory denoted by a sequence of measurement outcomes $\{a_1,\cdots,a_m\}$, the state of the target system is steered to
\begin{equation}\label{Eq:StateChangeFinal}
  \rho'=\frac{\M_{a_m}\cdots \M_{a_2}\M_{a_1}(\rho)}{p(a_1,a_2,\dots,a_m)},
\end{equation}
where $p(a_1,a_2,\dots,a_m)=\Tr[\M_{a_m}\cdots \M_{a_2}\M_{a_1}(\rho)]$ is the probability for this trajectory.

To estimate $\{v_k\}_{k=1}^s$, we arrange all the trajectories into new categories by defining a stochastic variable $\xi$ as a function of all the measurement outcomes $a_1,a_2,\dots a_m$. We can have many choices for the variable $\xi$, of which two typical cases are those in the repetitive and adaptive schemes. An important result in this paper is that for both schemes we provide a unifying expression for the probability distribution of $\xi$,
\begin{equation}\label{ker}
  p[\xi(a_1,a_2,\dots,a_m)]=\sum_{k=1}^s \Tr(P_k\rho)K(\xi,v_k),
\end{equation}
where $K(\xi,v)$ is a kernel function satisfying $\sum_{\xi}K(\xi,v)=1$ with $\xi$ being a discrete variable for finite $m$.

For the repetitive QPE scheme, each RIM cycle induces the same quantum channel $\Phi$ on the target system [see Fig. \ref{fig1}{\color{blue}(c)}]. The variable $\xi$ is chosen as the average results of all $m$ measurement results, $\xi=\sum_{i=1}^m (1-a_i)/m$, which coincides with $f_0$ in $F=(f_0,f_1)$ defined above \eqref{Fc}. In Sec. \ref{Sec:Repetitive}, we will show that the probability distribution of $F$ is composed of multiple binomial distributions
\begin{equation}
  p(f_0)\approx \sum_{k=1}^s {\rm Tr}(P_k\rho)e^{-mS(F\|F_k)},
\end{equation}
where $F_k=(p_{0k}, p_{1k})$ with $p_{\alpha k}=[1-(-1)^{\alpha}\cos(2v_k\tau+\phi)]/2$. The total probe time for a single sample of the repetitive scheme is $t=m\tau$. So for large $m$, the kernel for repetitive QPE is approximately a Gaussian kernel, which limits the estimation accuracy of repetitive QPE to the SQL ($\propto t^{-1/2}$).

The adaptive QPE scheme can further improve the estimation accuracy to the HL [Fig. \ref{fig1}{\color{blue}(e)}]. The basic idea is to represent the rescaled eigenvalues with binary expansions, $\vk=v_k/\|V\|=\sum_{i=1}^m {v_{k,i}}/{2^{m-i+1}}=0.v_{k,m}\cdots v_{k,2}v_{k,1}\cdots$ with $v_{k,i}\in\{0,1\}$ \footnote{We use the form of $v=0.v_1\dots v_m$ to be consistent with $a=0.a_m\dots a_2 a_1$ representing sequential measurement outcomes.}, and estimate the dominant $m$ bits of $v_k$ by $m$ sequential adaptive RIMs from the least significant digit (LSD) to most significant digit (MSD). Specifically, the $i$th bit of $v_k$ is estimated by the $i$th RIM with the free evolution time being $\tau_i=2^{m-i}\pi/\|V\|$ and the phase angle of $R_{\phi_i}(\pi/2)$ being $\phi_i=\pi-2\pi0.0a_{i-1}\cdots a_1$. Note that $\phi_i$ depends on all the prior $i-1$ measurement results. Then we choose the variable $\xi$ with the binary expansion, $\xi=a=0.a_m...a_2a_1$, whose distribution is found in Sec. \ref{Sec:adaptive} to be
\begin{equation}\label{pada}
  p(a)\approx \sum_{k=1}^s {\rm Tr}(P_k\rho)F_{2^{m}}(a-\vk)
\end{equation}
where $F_{N}(x):=\left[\frac{\sin(N\pi x)}{N\sin(\pi x)}\right]^2$ is a distorted Fej\'{e}r kernel. The total probe time for a single sample of the adaptive scheme is $t=\sum_{i=1}^m \tau_i=(2^m-1)\pi/\|V\|$. Since the Fej\'{e}r kernel $F_{2^m}(x)$ has the half-width proportional to $2^{-m}$, the estimation accuracy can reach the HL ($\propto t^{-1}$).


\subsection{Related works and our contributions}
To elucidate our contributions, we review efforts in related works about QPE algorithms for single and multiple eigenvalues.

For QPE of single eigenvalues with the target system prepared in the corresponding eigenstates, the ancilla measurements do not affect the input state of the target system, and many efforts have been dedicated to improving the estimation precision. In particular, adaptive Bayesian QPE algorithms have been developed \cite{Higgins2007,Berry2009,Cappellaro2012,Said2011} and widely used to measure physical quantities in quantum sensing, such as magnetic field strength \cite{Bonato2016, Zohar2023,Dinani2019} and hyperfine interaction strength \cite{Scerri2020}. Such Bayesian algorithms update the phase $\phi_i$ [Fig. \ref{fig1} {\color{blue}(e)}] to minimize the Holevo variance that quantifies the phase uncertainty, and can achieve the HL precision in single-eigenvalue QPE experiments \cite{Higgins2007,Berry2009}. The number of Bayesian updates can also be optimized to achieve the most accurate estimation with minimal resource consumption \cite{Berry2009,Cappellaro2012,Bonato2016}. More efficient Bayesian QPE based on rejection filtering has also been proposed \cite{Wiebe2016,Berg2021} and and experimentally demonstrated \cite{Paesani2017}. There are also non-adaptive algorithms that achieve the HL through postselection \cite{Higgins2009}. Recently a method based on quantum eigenvalue transformation of unitary matrices has been proposed to prepare the ground state and estimate the ground state energy through consecutive rotations of ancilla phase factors and controlled unitary gates \cite{Dong2022}.

There is emerging interest in developing efficient algorithms for multiple-eigenvalue QPE, especially for small-scale experiments. However, the methods for single-eigenvalue estimation in the above works cannot be naturally generalized to simultaneous estimation of multiple eigenvalues. Recently a Bayesian algorithm has been proposed for the simultaneous estimation of multiple arbitrary phases \cite{Gebhart2021}, with controlled unitary gates acting simultaneously on multiple copies of the target system prepared in different eigenstates. Other efforts proposed using only single-cycle measurements on a single ancilla, sampling on measurements with different evolution time and performing complex classical post-processing to realize multiple-eigenvalue QPE \cite{Somma2019,Dutkiewicz2022,Ding2023,Ding2023d}. Moreover, O'Brien \textit{et al.} developed a Bayesian QPE algorithm for multiple eigenvalues \cite{OBrien2019}, in which the information about the target system state in the sequential measurements is updated using the Bayesian theory. The basic idea in this latter work is quite similar to the adaptive scheme in this paper, however, due to lack of a unifying framework to deal with the measurement backaction in iterative QPE algorithms, this work did not rigorously analyze the state evolution of the target system and elucidate its role in multiple-eigenvalue QPE.

Some early works realized that the measurement backaction of QPE on the target system can be utilized to generate eigenstates \cite{Travaglione2001,Abrams1999}, by directly analyzing the state collapse of the target system for conventional QPE algorithms. In this paper, we provide a theoretical framework based on sequential quantum channels to systematically describe the measurement backaction in iterative QPE algorithms, and find that such backaction can help realize multiple-eigenvalue QPE. This framework is superior to the state collapse analysis in that the measurement backaction is treated as a quantum channel independent of the target system state, and also superior to the Bayesian theory in that it can account for both the statistics of ancilla measurements and its backaction on the target system. Therefore such a framework can provide a unifying picture for the repetitive and adaptive QPE schemes in this paper.



\section{Repetitive QPE scheme}\label{Sec:Repetitive}
In this section, we systematically investigate the backaction and statistics of sequential RIMs in the repetitive QPE scheme. By analyzing the asymptotic behavior of sequential quantum channels on the target system generated by RIMs, we find the backaction of repetitive measurements is steering the target system to the eigenspaces of $V$, corresponding to a PM on the target system. Then we analyze the measurement statistics by decomposing the average dynamics of repetitive quantum channels into stochastic trajectories. 
\subsection{Measurement backaction}
To analyze the backaction of sequential RIMs on the target system, it is illuminating to study the behaviors of sequential applications of the quantum channel $\Phi$. Previous works have studied the asymptotic behaviors of sequential quantum channels \cite{Albert2019,Burgarth2013a,Novotny2018,Blume-Kohout2010a}. Our recent work shows that sequential quantum channels with normal and commuting Kraus operators can simulate a PM in the asymptotic limit \cite{Ma2023}. The derivations below use this recent theoretical finding.

First we introduce the natural representation of quantum channels on the Hilbert-Schmidt (HS) space of the target system \cite{Bengtsson2006,Watrous2018}.
The space of operators on the Hilbert space of the target system form a linear vector space called the HS space. This can be seen by reshaping a matrix operator into a column vector, i.e., $X=\sum_{i,j=1}^d x_{ij}|i\rangle\langle j|\leftrightarrow |X\rangle\rangle=\sum_{i,j=1}^d x_{ij}|ij\rangle\rangle$, and defining the inner product in the HS space as $\langle\langle Y|X\rangle\rangle={\rm Tr}(Y^{\dagger}X)$, where $X$, $Y$ are operators on the target system. The superoperator $X(\cdot)Y$ is equivalent to a linear operator $X\otimes Y^{T}$ on the HS space with $Y^{T}$ being the transpose of $Y$,
so the channel $\Phi$ can be naturally represented as $\hat{\Phi}=\sum_{\alpha=0,1}\hat{\mathcal{M}}_{\alpha}$, where $\hat{\mathcal{M}}_{\alpha}=M_{\alpha}\otimes M_{\alpha}^*$ with $M_{\alpha}^*$ being the Hermitian conjugate of $M_{\alpha}$ (note that we add hats for operators on the HS space). With the HS space, the probability to get outcome $\alpha$ is $\langle\langle \mathbb{I}|\hat{\mathcal{M}}_{\alpha}|\rho\rangle\rangle={\rm Tr}(M_{\alpha}\rho M_{\alpha}^{\dagger})$.

The channel $\hat{\Phi}$ can be recast into a neat form if the Kraus operators are rewritten as \cite{Ma2023}
\begin{align}\label{Kraus}
\begin{bmatrix}
        M_0 \\ M_1
    \end{bmatrix}=
    \begin{bmatrix}
        {\tilde{\lambda}}_{01} \  &\cdots \  &{\tilde{\lambda}}_{0s} \\
        {\tilde{\lambda}}_{11} \  &\cdots \  &{\tilde{\lambda}}_{1s}
    \end{bmatrix}
    \begin{bmatrix}
        P_1 \\ \vdots \\ P_s
    \end{bmatrix},
\end{align}
where $\tilde{\lambda}_{\alpha k}=[e^{-i v_k\tau}-(-1)^{\alpha}e^{i(\phi+v_k\tau)}]/2$ is the $k$th eigenvalue of $M_{\alpha}$. Then $\tilde{\bm{\lambda}}_k=[\tilde{\lambda}_{0k},\tilde{\lambda}_{1k}]^T$ is a unit column vector in a two-dimensional complex vector space due to $\sum_{\alpha=0,1}M_{\alpha}^\dagger M_{\alpha} =\mathbb{I}$, and $\{\tilde{\bm{\lambda}}_{k}\}_{k=1}^{s}$ is a set of such unit vectors. Then $\hat{\Phi}$ becomes a diagonal operator on the HS space,
\begin{equation}\label{phi}
    \hat{\Phi}=\sum_{k,l=1}^{s}\langle\tilde{\bm{\lambda}}_{l},\tilde{\bm{\lambda}}_{k}\rangle P_k\otimes P_l,
\end{equation}
with the eigenvalues $\{\tilde{\bm{\lambda}}_{l},\tilde{\bm{\lambda}}_{k}\rangle\}_{k,l=1}^{s}$. Since $|\langle\tilde{\bm{\lambda}}_{l},\tilde{\bm{\lambda}}_{k}\rangle|\leq\langle\tilde{\bm{\lambda}}_{l},\tilde{\bm{\lambda}}_{l}\rangle\langle\tilde{\bm{\lambda}}_{k},\tilde{\bm{\lambda}}_{k}\rangle=1$ due to the Cauchy-Schwarz inequality \cite{Garcia2017}, all the eigenvalues of $\hat{\Phi}$ lie within the unit disk of the complex plane. The eigenvectors of $\hat{\Phi}$ with eigenvalue 1 are called fixed points \cite{Arias2002}, and those with eigenvalues $e^{i\varphi}$ ($\varphi\neq0$) are rotating points \cite{Albert2019}. The HS subspace spanned by the fixed points and rotating points are
called asymptotic subspace (also known as peripheral or attractor subspace). If any two unit vectors in $\{\tilde{\bm{\lambda}}_{k}\}_{k=1}^{s}$ are not parallel, then the asymptotic subspace contains only fixed points, which corresponds to all eigenstates of $V$.

With sequential applications of $\hat{\Phi}$, the projections to the asymptotic subspace remain unchanged, while the projections to the other eigenspaces gradually vanish \cite{Albert2019,Burgarth2013a,Novotny2018,Blume-Kohout2010a}. So for large $m$, $\hat{\Phi}^m\approx \sum_{k=1}^s \hat{\mathcal{P}}_k$, with $\hat{\mathcal{P}}_k=P_k\otimes P_k$ corresponding to the projection superoperator $P_k(\cdot)P_k$.
This implies that the backaction of sequential RIMs is approximately a PM on the target system \cite{Ma2023,Ma2018,Wang2023,Liu2017a,BhaktavatsalaRao2019a,Dasari2022a}, which project the target system to eigenspaces of $V$.

By decomposing sequential quantum channels, we can further show that each projector is approximately the summation of all $F$ in a certain range. The normal and commuting Kraus operators [Eq. (\ref{Kraus})] also allow an exact solution of the measurement statistics for sequential RIMs. Since $[\hat{\mathcal{M}}_{0}, \hat{\mathcal{M}}_{1}]=0$, we can expand $\hat{\Phi}^m$ according to the binomial theorem, $\hat{\Phi}^m=\sum_{f_0}\hat{\mathcal{M}}(f_0)$ with
\begin{equation}\label{}
    \hat{\mathcal{M}}(f_0)=\frac{m!}{(mf_0)!(mf_1)!}\hat{\mathcal{M}}_0^{mf_0}\hat{\mathcal{M}}_1^{mf_1}.
\end{equation}
With a reasoning similar to that in Eq. (\ref{Fc}), we get
\begin{equation}\label{Eq:qchannelF}
    \hat{\mathcal{M}}(f_0)\approx \sum_{k=1}^s e^{-mS(F\|F_k)}\hat{\mathcal{P}}_k,
\end{equation}
where $F_k=(p_{0k}, p_{1k})$ with $p_{\alpha k}=|\tilde{\lambda}_{\alpha k}|^2=[1-(-1)^{\alpha}\cos(2v_k\tau+\Delta \phi)]/2$. So if $\tilde{\bm{\lambda}}_k$ is regarded as a wavefunction [Eq. (\ref{wave})], $F_k$ is just the probability amplitude distribution. For large $m$, the Gaussian distribution is highly concentrated at $F_k$, then $\hat{\mathcal{P}}_k$ is related to a narrow interval of $F$ around $F_k$, which can also be used to realize selective PMs for purification of the target system state \cite{Ma2023}.

\subsection{Measurement statistics}
We analyze the measurement statistics by deriving the probability distribution of $f_0$ as
\begin{equation}\label{Fq}
    p(f_0)=\langle\langle\mathbb{I}|\hat{\mathcal{M}}(f_0)|\rho\rangle\rangle\approx \sum_{k=1}^s {\rm Tr}(P_k\rho)e^{-mS(F\|F_k)},
\end{equation}
where we have used $\langle\langle\mathbb{I}|\hat{\mathcal{P}}_k|\rho\rangle\rangle={\rm Tr}(P_k\rho)$.
Eq. (\ref{Fq}) represents a summation of at most $s$ different binomial distributions around $F_1,\cdots, F_s$ [Fig. \ref{fig1}{\color{blue}(d)}]. The weight of the $k$th binomial distribution is ${\rm Tr}(P_k\rho)$, that is, the projection of the initial target system $\rho$ on the $k$th eigenspace of $V$. {So if the target system starts from an eigenstate of $V$, we can still obtain the single binomial distribution [Eq. (\ref{Fc})] as in classical phase estimation.} Interestingly, the initial maximally mixed state of the target system is desirable for QPE of multiple eigenvalues, since then all the binomial distributions can appear, and integration of the measurement results for $k$th distribution also heralds a selective PM $\hat{\mathcal{P}}_k$ on the target system \cite{Ma2023}.

Any two binomial distributions around $F_k$ and $F_l$ can are well separated if the distance between $F_k$ and $F_l$ is larger than the sum of the respective half widths of the distribution. This requires \cite{Ma2023}
\begin{equation}\label{Eq:mReqNoiseless}
  m> \frac{1}{2}|\ln \eta|\frac{(\sqrt{p_{0k}p_{1k}}+\sqrt{p_{0l}p_{1l}})^2}{(p_{0k}-p_{0l})^2},
\end{equation}
where $\eta$ is the ratio of the minimum hight to the maximum hight within the distribution width. Then for large $m$, all the binomial distributions can be well distinguished and therefore the eigenvalue set $\{v_k\}_{k=1}^s$ of the operator $V$ can be estimated within the SQL. Combined with DD control on the ancilla, this scheme can also be used to estimate the multiple eigenvalues of a unitary generated by time-varying operators (see Appendix \ref{Sec:QPEfForTimeVarying}).

However, to estimate all of the eigenvalues effectively, Eq. (\ref{Eq:mReqNoiseless}) indicates that the minimum repetition number of RIMs is $m\propto{1}/{(\Delta p)_{\min}^2}$ with $\Delta p=p_{0k}-p_{1l}$. If the spectrum of the operator $V$ is dense, this requirement can hardly be satified. In this case, the probability distribution of $f_0$ can still be used to estimate the response function \cite{Roggero2020, Hartse2023}, which is defined as 
\begin{equation}\label{}
  S(\omega)\approx\sum_{k=1}^s \langle\langle{P_k}| \hat{\Phi}^m|{\rho}\rangle\rangle \delta(\omega-v_k)=\sum_{k=1}^s \Tr(P_k\rho) \delta(\omega-v_k),
\end{equation}
where $\delta$ the Dirac delta function. Then the probability distribution in \eqref{Fq} can be rewritten as an integral transform (see Appendix \ref{Sec:appDensOpe} for details),
\begin{equation}\label{Eq:integral}
	    p(f_0)\approx\int S(\omega)K_{G}(f_0,\omega)\dd \omega,
\end{equation}
where $K_{G}(f_0,\omega)=\frac{1}{\sqrt{2\pi}m{\sigma(\omega)}}\exp{-\frac{[f_0-\mu(\omega)]^2}{2[\sigma(\omega)]^2}}$ is a Gaussian function, with the mean $\mu(\omega)=[1-\cos(2\omega\tau+\phi)]/2$ and the variance $\sigma(\omega)=\sqrt{\mu(\omega)[1-\mu(\omega)]/m}$ being functions of $\omega$.

\section{Adaptive QPE scheme}\label{Sec:adaptive}
Now we examine the backaction and statistics of sequential RIMs for the adaptive QPE scheme. We show the backaction of sequential RIMs is still steering the target system to eigenspaces of $V$. However, compared to that the repetitive QPE scheme, the steering process is accelerated by adaptive control, which allows us to postprocess the measurement results to reach HL-limited estimation of multiple eigenvalues. 

To illustrate the basic idea of adaptive QPE, we first approximate all the eigenvalues by the $m$-bit binary expansions, $\vk=v_k/\|V\|\approx \sum_{i=1}^m {v_{k,i}}/{2^{m+i-1}}=0.v_{k,m}\cdots v_{k,2}v_{k,1}$ with $v_{k,i}\in\{0,1\}$. If the initial state of the target system is an eigenstate with eigenvalue $v_k$, the probability to obtain outcome 0 for the $i$th round of RIM is
\begin{equation}
  p_{0k,i}=\frac{1}{2}[1-\cos(2v_{k} \tau_i+\phi_i)],
\end{equation}
where $\tau_i$ is the free evolution time and $\phi_i$ the rotation angle of the $i$th round. Then we choose $\tau_i=2^{m-i}\pi/\|V\|$ {\color{red} \footnote{In practical experiments, one can also utilize prior knowledge of $V $ and select $\tau_i=2^{m-i}\pi/\tau_0$ with $\tau_0 > ||V||$  based on the bound of $V$, then we can get ${v_k}/{\tau_0}$ rather than ${v_k}/||V||$.}}, so
\begin{equation}
\begin{aligned}
	    2v_{k}t_i(\mathrm{mod} 2\pi) &=2\pi 0.v_{k,i}\dots v_{k,1}	\\
	    &=2\pi0.v_{k,i}+\phi'_i,
\end{aligned}
\end{equation}
with $\phi'_i=2\pi 0.0v_{k,i-1}\dots v_{k,1}$ in which all $v_{k,j}$ for $j\leq i-1$ has been obtained in previous $i-1$ measurements. Then with $\phi_i=\pi-\phi_i'$, we have
\begin{equation}
  p_{0k,i}=\frac{1}{2}[1-\cos(2\pi0.v_{k,i}+\pi)],
\end{equation}
thus if the $i$th outcome is 1(0), the $i$th bit of $\vk$ is 1(0). So after $m$ measurements, we can exactly estimate $v_k$. The total time of operation is $t=\sum_{i=1}^m 2^{m-i}\pi/\|V\|=(2^m-1)\pi/\|V\|$.

In practice, the numbers of bits in the binary expansions of $\{\vk\}_{k=1}^s$ are larger than $m$ or even infinite, i.e., $v_k/\|V\|=0.v_{k,m}\cdots v_{k,2}v_{k,1}\cdots$, and the initial state of the target system may be arbitrary. In previous works \cite{Giedke2006,Bonato2016}, the initial target system state is approximated as a prior probability distribution, which is then updated according to Bayes' rule during sequential ancilla measurements. Such a treatment cannot accurately account for the measurement backaction from the ancilla to the target and also make it ambiguous to realize QPE of multiple eigenvalues.

For adaptive QPE, we can clarify this ambiguity by using the formalism of sequential quantum channels in Sec. \ref{Sec:Repetitive}, with the only difference being that the channel for the $i$th RIM depends on all previous measurement results. We can analyze the measurement backaction by decomposing sequential channels as $\hat{\Phi}_m\cdots \hat{\Phi}_2\hat{\Phi}_1=\sum_{a}\hat{\mathcal{M}}(a)$
with
\begin{equation}
  \hat{\mathcal{M}}(a)=\hat{\mathcal{M}}_{a_m}\cdots \hat{\mathcal{M}}_{a_2}\hat{\mathcal{M}}_{a_1},
\end{equation}
where $a=0.a_m\dots a_2 a_1$ corresponds to the sequence of measurement outcomes $\{a_1,a_2,\cdots,a_m\}$. An interesting result in this paper is that we obtain the following exact formula for $\hat{\mathcal{M}}(a)$ (see Appendix \ref{subSec:deriveMMa} for the derivation),
\begin{equation}\label{Eq:Mma2}
  \hat{\mathcal{M}}(a)=\sum_{k,l=1}^s\sqrt{F_{2^m}(a-\vk)F_{2^m}(a-\vv_l)}P_k\otimes P_l,
\end{equation}
where $F_{2^m}(x)$ is the distorted Fej\'{e}r kernel as defined below Eq. \eqref{pada}, wit a distribution width proportional to $2^{-m}$. Since $v_k\neq v_l$ for $k\neq l$, for a relatively large $m$, $F_{2^m}(a-\vk)$ and $F_{2^m}(a-\vv_l)$ have negligible overlap and $\sqrt{F_{2^m}(a-\vv_k)F_{2^m}(a-\vv_l)}\to 0$. Then $\mm(a)$ can be approximated as
\begin{equation}\label{Eq:MMa}
	    \hat{\mathcal{M}}(a)\approx\sum_{k=1}^s F_{2^{m}}(a-\vk)\hat{\mathcal{P}}_k.
\end{equation}
Since $\sum_{a}F_{2^m}(a-\vk)=1 $, sequential adaptive quantum channels approximately induce a PM on the target system, $\hat{\Phi}_m\cdots \hat{\Phi}_2\hat{\Phi}_1\approx\sum_{k=1}^s\hat{\mathcal{P}}_k$, which is the same as the case of repetitive QPE. Then a specific projector $\hat{\mathcal{P}}_k$ corresponds to the integration of $\hat{\mathcal{M}}(a)$ over the range of $a$ near $\vk$.

For measurement statistics, the probability to get the measurement result $a$ is
\begin{equation}\label{Eq:Prob.Adap}
    p(a)=\langle\langle\mathbb{I}|\hat{\mathcal{M}}(a)|\rho\rangle\rangle\approx \sum_{k=1}^s {\rm Tr}(P_k\rho)F_{2^{m}}(a-\vk).
\end{equation}
Similar to Eq. \eqref{Fq}, the above formula represents a summation of at most $s$ distribution peaks around the set of rescaled eigenvalues $\{\vk\}$ with the weight of the $k$th distribution being $\Tr(P_k\rho)$. For dense spectra of $V$, Eq. \eqref{Eq:Prob.Adap} can be still be written in the form of Eq. \eqref{Eq:integral}, with the kernel there replaced by the Fej\'{e}r kernels. The difference is that the peaks determined by Fej\'{e}r kernels have the half-width proportional to $2^{-m}$, which indicates the error $\Delta\sim O(2^{-m})\propto t^{-1}$ obeys the HL. So the adaptive scheme is better than the repetitive scheme in resolving different eigenvalues and the response function.

\section{Examples}\label{Sec:Perf}

\begin{figure}
\includegraphics[width=3.5in]{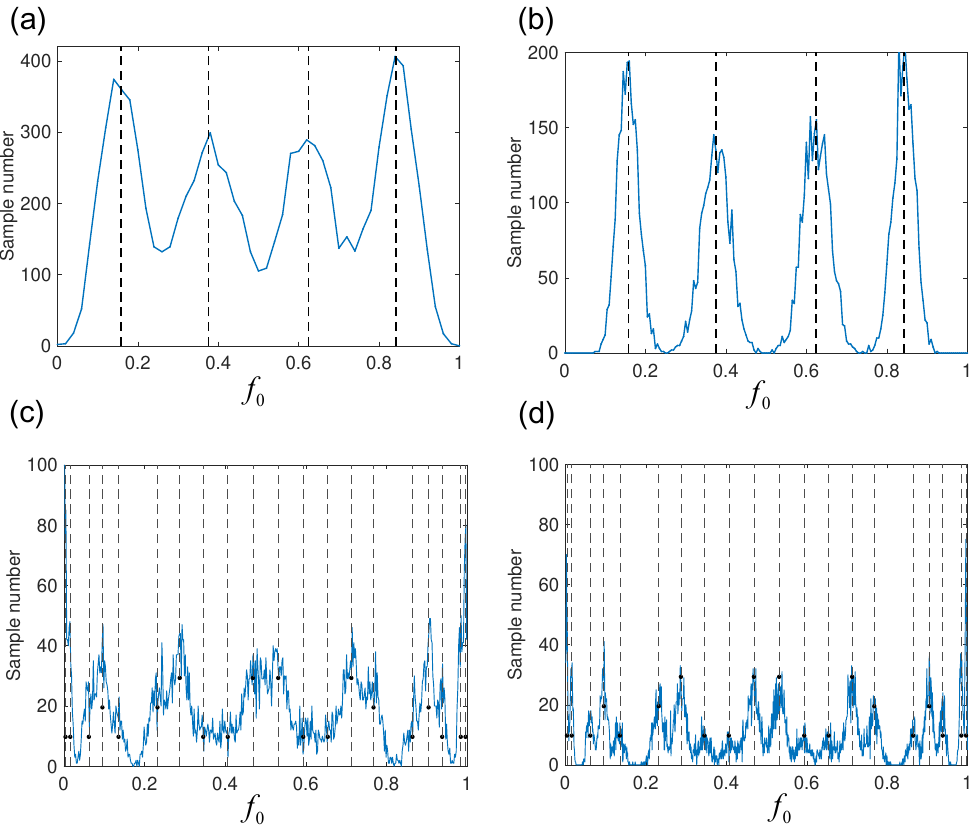}
\caption{ Monte Carlo simulations of repetitive QPE for a spin-star model. (a,b) For a target system containing two qubits, the number of samples as a function of $f_0$ for different repetition times of sequential
RIMs: (a) $m$ = 50, (b) $m$ = 200. The parameters are $A_1t=0.52$ rad, $A_2t=1.04$ rad. (c,d) is similar to (a,b) but for a target system containing five qubits: (c) $m$ = 500, (d) $m$ = 1000. The parameters are $[A_1t,A_2t,A_3t,A_4t,A_5t]=[ 0.13, 0.50, 0.38, 0.88,    1.01]$ rad. The black points in (a-d) represent the exact response function density (rescaled by the total sample number). The initial state of the target qubits is an equal superposition of all the eigenstates of $V$. All the simulations contain $10^4$ samples with $\phi=\pi/2$.}
\label{MCrep}
\end{figure}

\begin{figure}
\includegraphics[width=3.3in]{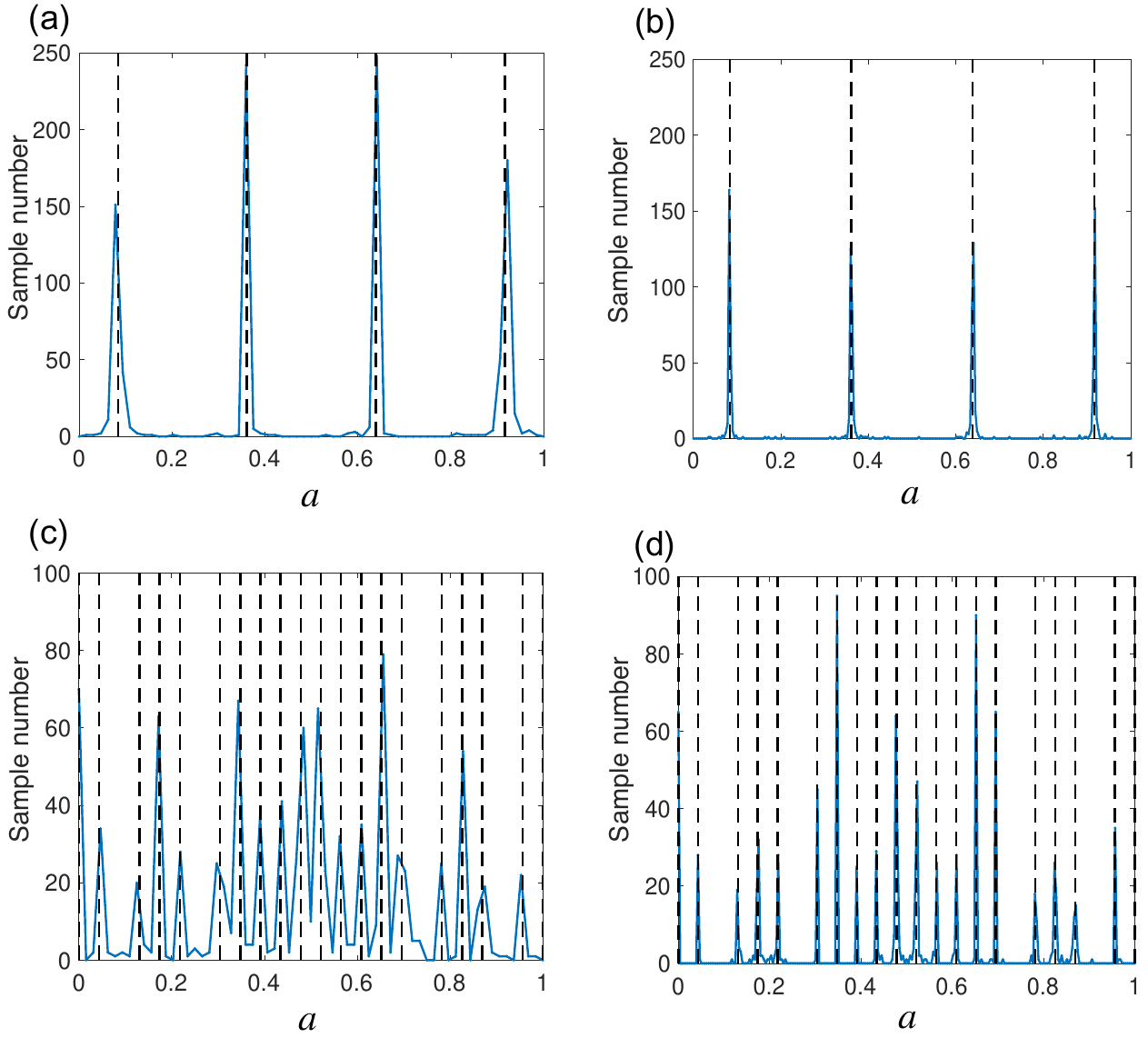}
\caption{ Monte Carlo simulations of adaptive QPE for a spin-star model. Similar to Fig. \ref{MCrep}, the target system contains two qubits in (a-b) and five qubits in (c-d). The numbers of adaptive RIMs are $m=6$ in (a,c) and $m=8$ in (b,d). The exact rescaled eigenvalues are indicated by black dashed lines, and
 the initial state of the target qubits is an equal superposition of all the eigenstates of $V$. All the simulations contain $10^3$ samples.}
\label{MCada}
\end{figure}

\begin{figure*}
\centering
  \includegraphics[width=18cm]{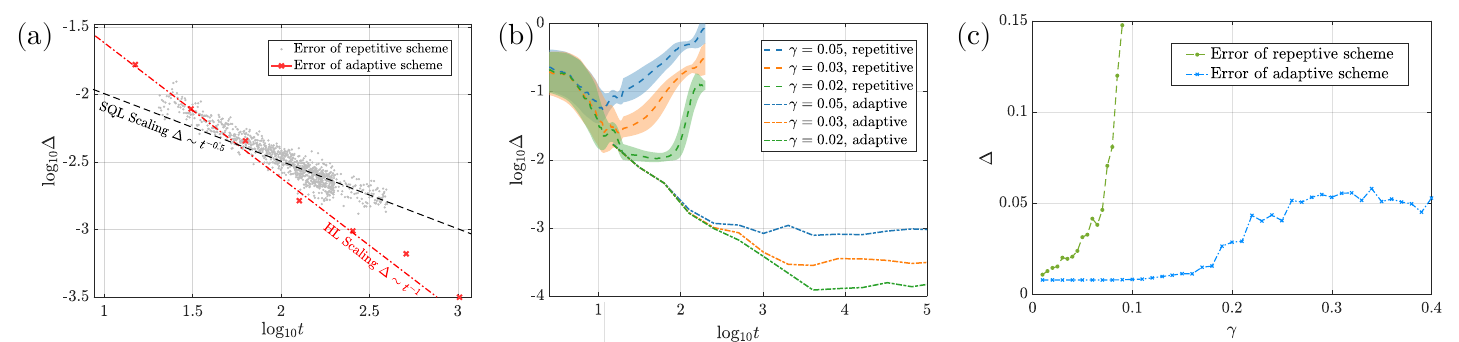}
  \caption{Comparison of performance and noise-resilience of the repetitive and adaptive QPE schemes. (a) The estimation error $\log_{10}\Delta$ of both schemes as a function of the total evolution time $t=m\tau$ for a noiseless target system. The black dashed and red dotted lines represent the SQL and HL respectively. (b) is similar to (a) but for a noisy target system with different noise strengths $\gamma=0.02,$ 0.03 and 0.05. For the repetitive scheme (dashed lines), the data is obtained by taking a moving average of the mean error of 10 identical experiments, with 90\% confidence intervals indicated by shading. For adaptive scheme (dash-dot lines), the data is the mean error of 10 identical experiments. (c) The estimation error $\Delta$ as a function of the noise strength $\gamma$. The error of the repetitive scheme (green dashed line) increases dramatically with $\gamma$ grows while the growth of that of the adaptive scheme (blue dash-dot line) is much slower. A theoretical bound is indicated by a black dashed line. In all the simulations, we take $5\times 10^3$ samples for the repetitive scheme and $10^3$ samples for the adaptive scheme. In the repetitive scheme, we take $\phi=\pi/2$, $\tau=0.2$ for every round of RIM.}\label{Fig:Perf}
\end{figure*}

Finally we compare the performance and noise resilience of the two QPE schemes for a concrete model. We consider a spin-star model describing an ancilla qubit inhomogeneously coupled with $K$ target qubits \footnote{We take the transformation $\sum_{j=1}^K \bm{\sigma}_{j}\cdot \mathbf{n}_j\to\sum_{j=1}^K\bm{\sigma}_{j}\cdot \mathbf{n}_j+\mathbb{I}_j$ to ensure the eigenvalues are positive, which is suitable with adaptive scheme. The spectrum of $\sum_{j=1}^K\bm{\sigma}_{j}\cdot \mathbf{n}_j\to\bm{\sigma}_{j}$ is $v_k'=\sum_{j=1}^K \eta'_{jk} A_j/4$ for $\eta'_{jk}\in\{-1,1\}$, which is symmetric with respect to 0. Adaptive scheme can deal with this type of spectrum with an additional procedure of post-processing, see Appendix \ref{App:nega}},
\begin{equation}\label{Hstar}
    H_{\rm star}=\sigma_q^z\otimes\sum_{j=1}^K \frac{A_j}{4}(\bm{\sigma}_{j}\cdot \mathbf{n}_j+\mathbb{I}_j),
\end{equation}
where $\bm{\sigma}_{j}=(\sigma_{j}^x, \sigma_{j}^y, \sigma_{j}^z)$ and $\mathbb{I}_j$ is the Pauli vector and identity operator for the $j$th target spin, and $\mathbf{n}_j=(n_j^x,n_j^y,n_j^z)$ is a unit vector. The operator $V=\sum_{j=1}^K A_j(\bm{\sigma}_{j}\cdot \mathbf{n}_j+\mathbb{I}_j)$ has eigenvalues $v_k=\sum_{j=1}^K \eta_{jk} A_j/2$ for $\eta_{jk}\in\{0,1\}$, so we can use QPE to estimate $\{v_k\}_{k=1}^{2^K}$ and thus the coupling strengths $\{A_j\}_{j=1}^K$. To describe the accuracy in estimating the eigenvalues, we use the mean absolute error
\begin{equation}
  \Delta=\frac{1}{2^K}\sum_{k=1}^{2^K} \abs{\tilde v_k-v_k},
\end{equation}
where $\{\tilde v_k\}$ is the set of estimated eigenvalues.

For a noiseless target system with the Hamiltonian in Eq. \eqref{Hstar}, we perform Monte Carlo simulations to simulate both repetitive and adaptive QPE schemes.
For repetitive QPE, we show the measurement statistics of repetitive RIMs for two target spins in Figs. \ref{MCrep}{\color{blue}(a)} and \ref{MCrep}{\color{blue}(b)} and five target spins in Figs. \ref{MCrep}{\color{blue}(c)} and \ref{MCrep}{\color{blue}(d)}. In the former case, one can see four distinct binomial distributions as the repetition number $m$ of RIMs increases, corresponding to the four different eigenvalues of the operator $V$, while in the latter case, the number of binomial distributions becomes so large that different neighboring peaks cannot be well distinguished for a finite $m$. For adaptive QPE, we run similar simulations about both two-qubit and five-qubit target systems, and all the eigenvalues can be well estimated within $6\sim 8$ measurements (see Fig. \ref{MCada}). For both schemes we can well estimate the spectral density function of the target system, and the accuracy of repetitive QPE reaches SQL, while that of the adaptive QPE reaches HL [see Fig. \ref{Fig:Perf}(a)].

In practice, various noise terms are often unavoidable. We add to the ideal Hamiltonian in Eq. \eqref{Hstar} a noise operator $C=\sum_{j=1}^K\omega_j(\sigma_{j}^z+\mathbb{I}_j)/2$, which is a free Hamiltonian of the target system. For the repetitive QPE scheme, the condition for the repetitive QPE to accurately estimate eigenvalues relies on both $m$ and $\gamma$. If the intensity of noise is large, our recent work shows that the target system will be quickly depolarized to the maximally mixed state in sequential RIMs as the number $m$ of sequential RIMs increases \cite{Qiu2024}, which makes it impossible to estimate the eigenvalues. However, the repetitive scheme can still work with a weak noise, which can be interpreted by the theory of metastability \cite{Jin2024}. The theory shows that if $\gamma=||C||/||V||\ll 1$, the target system can still be polarized to extreme metastable states (EMSs) $\{\rho_v\}_{v=1}^s$, which are the eigenstates of $V$ up to some correlations in metastable region of $m$ (see Appendix \ref{Metastability} for technical details). For the adaptive QPE scheme, the quantum channels and the Kraus operators are different for different rounds of RIMs, so it is impossible to predict the behaviors of sequential quantum channels by the spectral analysis. However, we can still directly simulate the measurement statistics and backaction of the specified sequence of outcomes.


The simulation results for the target system with such a coherent noise (typically $\gamma\sim 10^{-2}$) is shown in Fig. \ref{Fig:Perf}(b). For the repetitive scheme, the metastable region of $m$ for accurate estimation is narrow, and increasing $m$ further only results in larger errors [see dashed lines in Fig. \ref{Fig:Perf}(b)]. For the adaptive scheme, the depolarization process of the target system can be largely suppressed, and the estimation error $\Delta$ first obeys the HL as a function of $m$ and then saturates as $m$ increases further [see dash-dot lines in Fig. \ref{Fig:Perf}(b)]. We also compare the estimation errors of two schems for fixed intensity of noise, and clearly demonstrates that the adaptive scheme is more immune to coherent noise [see Fig. \ref{Fig:Perf}(c)].

Apart from the coherent noise, the target system may also suffers from incoherent noise described by Markovian master equations. We numerically simulate the effect of such incoherent noise on the performance of the two QPE schemes in Appendix \ref{incoherent}, and find that the adaptive scheme is also more noise resilient than the repetitive scheme.


\section{Conclusions and outlooks}
We have developed a channel-based framework to realize iterative QPE of multiple eigenvalues based on sequential RIMs of an ancilla qubit without state preparation. We thoroughly investigate two specific QPE schemes with this framework, including the repetitive scheme and the adaptive scheme. In the repetitive scheme, sequential measurements probabilistically project the target system to different eigenstates of the estimated unitary operator. The measurement statistics can display multiple distribution peaks, which can be used to precisely estimate the eigenvalues of the operator on a target quantum system within the SQL. While for the adaptive scheme, sequential adaptive measurements can more efficiently project the system to eigenstates compared to the repetitive scheme, and the measurement statistics show the pattern of multiple Fej\'{e}r kernels, so that we can realize QPE within the HL. We also analyze the performance and noise resilience of both schemes for a specific spin-star model. Compared to the repetitive scheme, the adaptive scheme is more accurate and noise-resilient, at the expense of more advanced feedback control of the ancilla.

Our work lays a solid theoretical foundation for multi-eigenvalue QPE algorithms. It will be interesting future works to use the channel-based framework to explore various superior traditional QPE algorithms or further refine the adaptive techniques to harness even greater efficiencies and accuracies. The QPE schemes in this work also shed light on multi-parameter quantum sensing with traditional RIMs.

We add some comments about the practical implementation of two QPE schemes. Both schemes in this paper are hardware-efficient since they require only an ancilla qubit. For the experimental platforms admitting adaptive control, the adaptive QPE scheme is superior to the repetitive one in both estimation accuracy and noise resilience. However, real-time adaptive control, requiring high-fidelity single-shot readout (or PM) of the ancilla and fast classical data processing, has only be realized in a few systems, such as NV centers at low temperatures \cite{Robledo2011,Blok2014,Bonato2016} and superconducting circuits \cite{Sayrin2011}. In comparison, we expect that the repetitive QPE scheme is easy to realize in most platforms as it requires no adaptive control.

\section{Acknowledgement}
We thank Ren-Bao Liu and Cristian Bonato for very helpful discussions. The research is supported by the National Natural Science Foundation of China (No. 12174379, No. E31Q02BG), the Chinese Academy of Sciences (No. E0SEBB11, No. E27RBB11), the Innovation Program for Quantum Science and Technology (No. 2021ZD0302300) and Chinese Academy of Sciences Project for Young Scientists in Basic Research (YSBR-090).

\appendix
\begin{widetext}

\section{Asymptotic behavior of sequential quantum channels generated by iterative QPE}\label{App:Sequential channel}
Here we study the sequential quantum channels induced by iterative QPE with a general sequences. We write Eq. \eqref{Kraus} in HS space and consider $\tau$ and $\phi$ as variables labeling $\tau_i$ and $\phi_i$, then we obtain the Kraus operators of a general RIM cycle,
\begin{equation}
\begin{aligned}
	     \mm_0(\tau_i,\phi_i)&=\sum_{k,l}[e^{-i(v_k-v_l)\tau_i}-e^{i((v_l+v_k)\tau_i+\phi_i)}-e^{-i((v_l+v_k)\tau_i+\phi_i)}+e^{i(v_k-v_l)\tau_i}]P_k\otimes P_l/4	\\
	     &=\sum_{k,l}\{\cos[(v_k-v_l)\tau_i]-\cos[(v_k+v_l)\tau_i+\phi_i]\}P_k\otimes P_l/2,
\end{aligned}
\end{equation}
and
\begin{equation}
 \mm_1(\tau_i,\phi_i) =\sum_{k,l}\{\cos[(v_k-v_l)\tau_i]+\cos[(v_k+v_l)\tau_i+\phi_i]\}P_k\otimes P_l/2.
\end{equation}
Then the single quantum channel is
\begin{equation}
  \phii(\tau_i,\phi_i)=\phii(\tau_i)=\sum_{k,l}\cos[(v_k-v_l)\tau_i]P_k\otimes P_l,
\end{equation}
and the sequential quantum channel is
\begin{equation}
  \phii_{\rm tot}=\phii(\tau_m)...\phii(\tau_2)\phii(\tau_1)=\sum_{k,l}\prod_{i=1}^m\cos[(v_k-v_l)\tau_i]P_k\otimes P_l.
\end{equation}
We don't take degenerations into account, so $\cos[(v_k-v_l)\tau_i]<1$, for the asymptotic behavior, i.e. $m$ is relatively large, $\prod_{i=1}^m\cos[(v_k-v_l)\tau_i]=\delta_{k,l}$, then
\begin{equation}
  \phii_{\rm tot}=\sum_{k=1}^s P_k\otimes P_k,
\end{equation}
which is the summation of projectors. So by choosing the sequence $\{\tau_i\}_{i=1}^m$ properly, QPE and target state purification (projection) can be realized fast.

\section{QPE for a unitary generated by a time-dependent operator}\label{Sec:QPEfForTimeVarying}
So far we have used sequential RIMs to realize QPE of the eigenvalues $\{v_k\}_{k=1}^s$ of a static operator $V$, which is the generator of the unitary operator to be estimated. When combined with DD control of the ancilla, such a scheme can also estimate the eigenvalues of a unitary operator generated by a time-varying quantum operator $V(t)$.

The basis idea is similar to that of classical ac signal sensing \cite{Kotler2011, DeLange2011}. For a time-dependent quantity $v(t)=\sum_{\omega}v(\omega)e^{i\omega t}$ with $v(\omega)=v(-\omega)$, the RIM scheme can detect the integration of such a quantity $\int_0^t v(t')dt'$. With DD control of the ancilla (consisting of a sequence of $\pi$ flips at times $\{t_1,t_2,\cdots,t_N\}$ for the ancilla evolution from 0 to $t$ during the RIM cycle), the accumulated quantity becomes \begin{equation}
  \int_0^t f(t')v(t')dt'=\sum_{\omega}v(\omega)F(\omega,t),
\end{equation}
where $f(t)$ is the DD modulation function jumping between $+1$ and $-1$ every time the ancilla is flipped by a DD pulse and $F(\omega,t)=\int_0^t f(t')e^{i\omega t}dt'$ is the DD filter function. For $v(t)$ composed of a single or several harmonic tones, the amplitude $v(\pm\omega)$ of a signal tone can be detected if $f(t)$ is periodic and its frequency matches that of the tone so that $F(\omega,t)$ only filters such a tone.

For a time-dependent operator $V(t)$, $U_{\alpha}=\mathcal{T}e^{-i(-1)^{\alpha}\int_0^t V(t')dt'}$, with $\mathcal{T}$ denoting the time-ordering. We find that the RIM scheme can estimate the eigenvalues of the unitary operator $U_{\alpha}$ with the condition $[U_0, U_1]=0$. This condition can be met when $U_0$ and $U_1$ are well approximated by the first-order Magnus expansion, i.e., $U_0\approx e^{-i\int_0^t V(t')dt'}\approx U_1^{\dagger}$, then the RIM scheme can estimate the eigenvalues of the operator $\int_0^t V(t')dt'$. Specifically, we consider a model with the Hamiltonian
\begin{equation}\label{}
    H=f(t)\sigma_q^z\otimes V+\mathbb{I}_q\otimes H_0,
\end{equation}
where $H_0=\sum_{i=1}^d \varepsilon_i|i\rangle\langle i|$ is the free Hamiltonian, and $V=\sum_{i,j=1}^d V_{ij}|i\rangle\langle j|=\sum_{\omega}V(\omega)$ with $V(\omega)=V(-\omega)^{\dagger}=\sum_{\varepsilon_i-\varepsilon_j=\omega}V_{ij}|i\rangle\langle j|$. If $[V,H_0]=0$, it is still static and thus the above RIM scheme works. However, if $[V,H_0]\neq0$, the operator $V$ becomes $V(t)=e^{iH_0t}Ve^{-iH_0t}=\sum_{\omega}e^{i\omega t}V(\omega)$ in the interaction picture, which can be termed as time-varying operator. Then DD control of the ancilla can select the component $V(\pm\omega)$ and filter out the other parts in $V$ through the filter function $F(\omega,t)$, if the frequency of $f(t)$ matches $\omega$. Then we have 
\begin{equation}\label{}
    U_1\approx U_0^{\dagger}\approx{\rm exp}\{-i[F(\omega,t)V(\omega)+F^*(\omega,t)V^{\dagger}(\omega)])\},
\end{equation}
whose eigenvalues can be estimated as in the case for QPE of a static operator $V$.

As an example, we consider the spin-star model with an external magnetic field,
\begin{equation}\label{Hstar2}
    H_{\rm star}=\sigma_q^z\otimes\sum_{j=1}^K \frac{A_j}{4}(\bm{\sigma}_{j}\cdot \mathbf{n}_j)+\frac{\omega}{2} \sum_{j=1}^K\sigma_{j}^z,
\end{equation}
where the first term denotes the inhomogeneous coupling between the ancilla and $K$ target spins, and the second term is an homogeneous Zeeman term of the target spins induced by an external magnetic field.
If $\omega=0$, the Hamiltonian returns to Eq. (\ref{Hstar}) and the eigenvalues of $V$ can be estimated without DD control. With the Zeeman term ($\omega\neq0$), we can apply the $N$-pulse Carr-Purcell-Meiboom-Gill (CPMG) control to the ancilla with the flip pulses at times $\{t_l=(2l-1)\tau\}_{l=1}^N$ ($2\tau=t/N$).
With the resonant DD condition ($2\tau=\pi/\omega$),
 \begin{equation}
  U_1\approx U_0^{\dagger}\approx \prod_{j=1}^K{\rm exp}\left[-\frac{iA_{j}^{\perp}N}{4\omega}(\cos\kappa_j\sigma_{j}^{x}-\sin\kappa_j\sigma_{j}^{y})\right],
\end{equation}
with $A_j^{\perp}=A_j[(n_j^x)^2+(n_j^y)^2]$ and $\kappa_j={\rm arg}(F(\omega,t))+{\rm arctan}(A_j^y/A_j^x)$, so by sequential RIMs we can estimate $\{A_j^{\perp}\}$, which are the components of the interaction fields $\{A_j\mathbf{n}_j\}$ perpendicular to the external magnetic field. With several different directions of the external magnetic field, we can estimate both $\{A_j\}$ and $\{\mathbf{n}_j\}$ \cite{Ma2016a}. We perform Monte Carlo simulation for this circumstance, the information of $\{A_j^{\perp}\}$ can be obtained by analyzing the peaks.


\begin{figure*}
\centering
  \includegraphics[width=18cm]{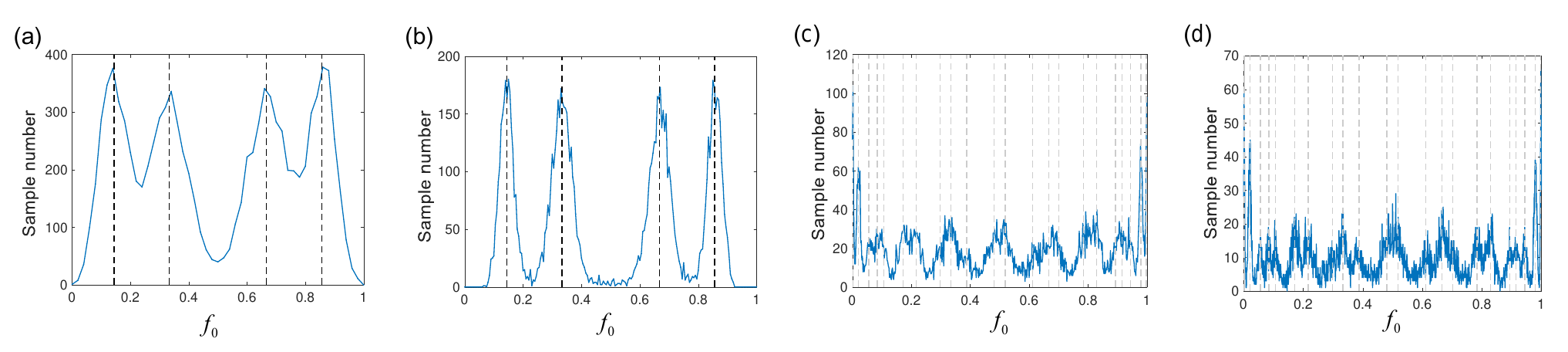}
  \caption{Monte Carlo simulation of QPE in a spin-star model containing Zeeman term. (a-b) For the model in Eq. (\ref{Hstar2}) with $\omega/A_1=2\omega/A_2=10$, $\mathbf{n}_1=(0.50,0.50,0.71)$, $\mathbf{n}_2=(0.10,0.10,0.99)$ and the CPMG pulse number $N=4$, the histogram of the number of samples as a function of the frequency $f_0$ for different repetition times in sequential RIMs: (a) $m$ = 50, (b) $m$ = 200. (c-d) are similar with (a-b), the parameters are $\omega/[A_1,A_2,A_3,A_4,A_5]=[30,6,10,12,20]$ and $\mathbf{n}_j=(n_{j,x},n_{j,x},\sqrt{1-2n_{j,x}^2})$ with $[n_{1,x},n_{2,x},n_{3,x},n_{4,x},n_{5,x}]=[0.5,0.1,0.3,0.7,0.5]$, $m=200$ in (c) and $m=500$ in (d). The dashed lines in (a-d) represent the exact eigenvalues. The initial state of the target qubits are an equal superposition of all the four eigenstates of $V$. All the simulations contain $10^4$ samples with $\phi=\pi/2$. }
\end{figure*}

\section{Details about the spectral-density operator and response function}\label{Sec:appDensOpe}
For an observable $V=\sum_{k=1}^s v_k P_k$ of a quantum system, we can define its spectral-density operator as \cite{Roggero2020,Hartse2023}
\begin{equation}
  \varrho(\omega)=\delta (\omega-V)=\sum_{k=1}^s \delta (\omega-v_k)P_k,
\end{equation}
where $\delta(\omega)$ is the Dirac delta function satisfying $\int \delta(\omega)d\omega=1$. Given a density matrix $\rho$ of the system, we can define its response function as
\begin{equation}
   S(\omega)=\Tr[\varrho(\omega)\rho]=\sum_{k=1}^s\Tr(P_k\rho)\delta (\omega-v_k).
\end{equation}

For the target system in our paper, given an initial state $\kett\rho$ and the repetitive number $m$ of RIMs, the final state is $\kett{\rho'}=\phii^m\kett\rho$. Then we investigate the response function
\begin{equation}
\begin{aligned}
	    S(\omega)&=\Tr[\varrho(\omega)\rho']
        =\langle\langle{\varrho(\omega)}|\hat{\Phi}^m|{\rho}\rangle\rangle
        =\sum_{k=1}^s \langle\langle{P_k}|\hat{\Phi}^m|{\rho}\rangle\rangle \delta(\omega-v_k).
\end{aligned}
\end{equation}
Since the natural representation of a quantum channel is a linear operator on the HS space, it can be spectrally decomposed as \cite{Wolf2010}
\begin{equation}
  \hat\Phi=\sum_i \lambda_i |R_i\rangle\rangle \langle\langle L_i|,
\end{equation}
where $\lambda_i=\abs{\lambda_i}\ee^{i\varphi_i}$ is the $i$th eigenvalue and $\kett{R_i}$($\kett{L_i}$) is the corresponding right (left) eigenvector, satisfying $\hat \Phi\kett{R_i}=\lambda_i \kett{R_i}$, $\hat \Phi^\dagger\kett{L_i}=\lambda_i^{*} \kett{L_i}$, and the biorthonormalization condition $\brakett{L_i}{R_j}=\mathrm{Tr}(L_i^\dagger R_j)=\delta_{ij}$. Then when $m$ is large enough, the contribution of decaying points (eigenspaces with $\abs{\lambda_i}<1$) can be omitted, and
\begin{equation}\label{phirho}
  \phii^m\kett\rho\approx \sum_{k=1}^s \kett{P_k}\brakett{P_k}{\rho}=\sum_{k=1}^s \Tr(P_k\rho)\kett{P_k}
\end{equation}
here we note that for sequential RIMs with Hamiltonian $H=\sigma_q^z\otimes V$, the fixed points of quantum channel are the projectors $\{|P_k\rangle\rangle\}$, and $\kett{R_i}=\kett{L_i}$ for fixed points.
So, the form of response function can be approximated as
\begin{equation}
  S(\omega)\approx\sum_{k=1}^s \Tr(P_k\rho)\delta(\omega-v_k).
\end{equation}


For large $m$, the probability distribution of $f_0$ is
\begin{equation}
\begin{aligned}
       p(f_0)\approx\sum_{k=1}^s\Tr(P_k\rho)\frac{1}{\sqrt{2\pi}m{\sigma_k}}\exp{-\frac{(f_0-p_{k0})^2}{2\sigma_k^2}},
\end{aligned}
\end{equation}
where $\sigma_k=\sqrt{p_{k0}p_{k1}/m}$. So $p({f_0})$ can be rewritten in the form of integral transformation,
\begin{equation}
\begin{aligned}\label{Hamiltonian}
	    p(f_0)\approx\int S(\omega)K_{G}(f_0,\omega)
	    = \sum_{k=1}^s \Tr(P_k\rho) K_{G}(f_0,v_k),	
\end{aligned}
\end{equation}
where $K_{G}(f_0,\omega)=\frac{1}{\sqrt{2\pi}m{\sigma(\omega)}}\exp{-\frac{[f_0-\mu(\omega)]^2}{2[\sigma(\omega)]^2}}$ is a Gaussian function, with the mean $\mu(\omega)=[1-\cos(2\omega\tau+\phi)]/2$ and the variance $\sigma(\omega)=\sqrt{\mu(\omega)[1-\mu(\omega)]/m}$ being functions of $\omega$.

\section{Fundamentals of metastability in repetitive RIMs}\label{Metastability}
In our previous works \cite{Jin2024,Qiu2024}, we point out that the fixed points of the channel in RIM depend on the commutativity of $V$ and $C$. If $[V,C]=0$, the fixed points must include the linear combinations of rank-one projections $\{|j\rangle\langle j|\}_{j=1}^s$; if $[V,C]\neq 0$, the fixed points must include the linear combinations of projection operators $\{\Pi_j\}_{j=1}^r$ ($r\leq s$), satisfying $\sum_{j=1}^r\Pi_j=\mathbb{I}$. Thus, for $[V,C]=0$, the asymptotic operation of sequential such channels is a polarizing channel (or a PM) on the target system, while for $[V, C]\neq0$ it can be a depolarizing channel at least for the subspace $\mathcal{H}_j$ with $\dim(\mathcal{H}_j)\geq2$.

Quantum metastability can happen when $C$ can be considered as a perturbation on $V$, i.e., $||C||\ll||V||$. Then the $s$-fold degeneration of fixed points breaks down, leaving $r$ fixed points $\{\Pi_j\}_{j=1}^r$ and the other $(s-r)$ decaying points with eigenvalue $|\lambda_j|\approx 1$ (called \textit{metastable points}). For such a channel $\hat\Phi$ with $r$ fixed points, and $s-r$ metastable points, quantum metastability can emerge. After sequentially applying the quantum channel for $m$ times, the target system state becomes
 {\begin{equation}
\begin{aligned}\label{3}
	    \hat\Phi^m\kett{\rho}&=\sum_{i=1}^r c_i\kett{\rho_{\rm fix}^i}+\sum_{j=r+1}^{s}c_j\lambda_j^m \kett{R_j}+\cdots\\
	    &\approx\sum_{i=1}^r c_i\kett{\rho_{\rm fix}^i}+\sum_{j=r+1}^{s}c_j\ee^{ m(\ln{\abs{\lambda_j}+\ii \varphi_j)}} \kett{R_j},
\end{aligned}
\end{equation}}
with $\lambda_j=\abs{\lambda_j}\ee^{\ii\varphi_j}$ and $c_j=\brakett{L_j}{\rho}={\Tr}(L_j^\dagger \rho)$. The contribution of the other decaying points decays fast as $m$ grows, and can be omitted when $m\gg \mu''=1/\abs{\ln\abs{\lambda_{s+1}}}$, while the metastable points cannot be neglected when $m\ll \mu' = 1/\abs{\ln \abs{\lambda_s}}$. So $\mu'$ and $\mu''$ delimit a metastable region:
\begin{equation}\label{Eq:cond.repAp}
  \frac{1}{|\ln|\lambda_{s+1}||}\ll m \ll \frac{1}{|\ln|\lambda_{s}||},
\end{equation}
where the metastable points with real eigenvalues act like fixed points, and those with complex eigenvalues act like rotating points.

Without loss of generality, we consider only the metastable points with real eigenvalues. In metastable region, we have $e^{m\ln|\lambda|}\approx 1$, so the target system state becomes (for appropriate interval of $m$)
\begin{equation}
\begin{aligned}\label{channelmeta}
	    \hat\Phi^m\kett{\rho}&\approx\sum_{i=1}^r c_i \kett{\rho_{\rm fix}^i}+\sum_{j=r+1}^{s}c_j\kett{R_j}
        \approx \sum_{v=1}^{s}{p_v}\kett{\rho_v},
\end{aligned}
\end{equation}
where $c_i=\Tr(L_i\rho)$ and $\kett{\rho_{\rm fix}^i}$ is $i$th fixed point. The second line is a transformation of the first line, where $\{\rho_v\}$ is a set of disjoint extreme metastable states (EMSs) \cite{Gaveau2006, Macieszczak2016, Jin2024}, and $ p_v=\mathrm{Tr}( P_v\rho)$ satisfying $\sum_{v}  p_v=1$. Here $\{ P_v\}$ is a set of observables satisfying $\langle\langle{ P_v}|{\rho_u}\rangle\rangle=\delta_{vu}$, $ P_v\geq 0$ and $\sum_v { P_v}=\mathbb{I}$.

From Eq. (\ref{channelmeta}), we can see that in metastable region, the target system is projected to each EMS, and we have proved that the EMSs for the case of sequential RIMs are $ \{|j\rangle\langle j|\}_{j=1}^d$ up to some corrections \cite{Jin2024}, then $\phii^m \kett{\rho}\approx\kett{P_k}\brakett{P_k}{\rho}=\sum_{k=1}^s \Tr(P_k\rho)\kett{P_k}$, which is exactly Eq. (\ref{phirho}).
Beyond the metastability region ($m\gtrsim\mu'$), the contribution of the second term in Eq. (\ref{channelmeta}) decreases exponentially as $m$ increases, so the system gradually leaks from metastable states and depolarizes toward the stationary states corresponding to the fixed points. 

\section{Details about adaptive QPE}
\subsection{Details about backaction of sequential RIMs for adaptive QPE}\label{subSec:deriveMMa}
We provide details about the backaction of sequential RIMs on the target system for the adaptive QPE scheme, including the derivation of Eq. \eqref{Eq:Mma2} and Eq. \eqref{Eq:MMa} in Sec. \ref{Sec:adaptive}. Since $M_{a_i}=\sum_k[ e^{-iv_k\tau_i} -e^{i(\phi_i+v_k\tau_i-a_i\pi)}] P_k/2$, its corresponding superoperator is
\begin{equation}
\begin{aligned}
	    \mm_{a_i}&=M_{a_i}\otimes M_{a_i}^*	\\
	    &=\sum_{k=1}^s\sum_{l=1}^s\frac{\cos[(v_k-v_l)\tau_i]-\cos[(v_k+v_l)\tau_i+\phi_i-a_i\pi]}{2}P_k\otimes P_l.
\end{aligned}
\end{equation}
We choose $\phi_i=\pi-2\pi 0.0a_{i-1}\dots a_{1}$, so $\phi_i-a_i\pi=\pi-2\pi a\times 2^{m-i}$. So we have
\begin{equation}
\begin{aligned}
	    \hat{\mathcal{M}}(a)=&\hat{\mathcal{M}}_{a_m}\cdots \hat{\mathcal{M}}_{a_2}\hat{\mathcal{M}}_{a_1},	\\
	    =&\sum_{k,l=1}^s\prod_{i=1}^m\frac{\cos[(v_k-v_l)\tau_i]-\cos[(v_k+v_l)\tau_i+\phi_i-a_i\pi]}{2}P_k\otimes P_l\\
	    =&\sum_{k,l=1}^s\prod_{i=1}^m\frac{\cos[(\vv_k-a-(\vv_l-a))\times 2^{m-i}\pi]+\cos[(\vv_k+\vv_l)\times 2^{m-i}\pi-2\pi a\times 2^{m-i}]}{2}P_k\otimes P_l\\
	    =&\sum_{k,l=1}^s\prod_{j=0}^{m-1}{\cos[(a-\vv_k)\times 2^{j}\pi]\cos[(a-\vv_l)\times 2^{j}\pi]}P_k\otimes P_l.
\end{aligned}
\end{equation}
where we have changed the summation index by $j=m-i$ in deriving the last line. Then we consider the expression $\prod_{i=0}^{m-1} \cos^2 \left[(\vv_k-a) 2^{i}\pi\right]$,
\begin{equation}
  \begin{aligned}
	  &\prod_{i=0}^{m-1} \cos^2 \left[(a-\vv_k) 2^{i}\pi\right]
	    =\cos^2 \left[(a-\vv_k) 2^{m-1}\pi\right]\times \cos^2 \left[(a-\vv_k) 2^{m-2}\pi\right]\times \cdots \times \cos^2 \left[(a-\vv_k)\pi\right]\\
	    =&\left(\frac{\sin \left[(a-\vv_k) 2^{m}\pi\right]}{2\sin \left[(a-\vv_k) 2^{m-1}\pi\right]}\right)^2\times \left(\frac{\sin \left[(a-\vv_k) 2^{m-1}\pi\right]}{2\sin \left[(a-\vv_k) 2^{m-2}\pi\right]}\right)^2\times\cdots \left(\frac{\sin \left[(a-\vv_k) 2\pi\right]}{2\sin \left[(a-\vv_k)\pi\right]}\right)^2\\
	    =&\left(\frac{\sin \left[(a-\vv_k) 2^{m}\pi\right]}{2^m\sin \left[(a-\vv_k) \pi\right]}\right)^2
	    =F_{2^{m}}(a-\vv_k)	\\
\end{aligned}
\end{equation}
where $F_{N}(x):=\left[\frac{\sin(N\pi x)}{N\sin(\pi x)}\right]^2$ is the distorted Fej\'{e}r kernel defined in the main text. So $\hat{\mathcal{M}}(a)$ can be rewritten as
\begin{equation}\label{Eq:Mma2}
  \hat{\mathcal{M}}(a)=\sum_{k,l=1}^s\sqrt{F_{2^m}(a-\vv_k)F_{2^m}(a-\vv_l)}P_k\otimes P_l.
\end{equation}
Since $v_k\neq v_l$ (then $a_k\neq a_l$) for $k\neq l$, we have $\sqrt{F_{2^m}(a-\vv_k)F_{2^m}(a-\vv_l)}\to 0$ when $m$ is large enough so that $F_{2^m}(a-\vv_k)$ and $F_{2^m}(a-\vv_l)$ have negligible overlap. Then we get a approximate form of $\mm(a)$,
\begin{equation}
  \mm(a)\approx\sum_{k=1}^s F_{2^m}(a-\vv_k)\hat{\mathcal{P}}_k,
\end{equation}
which is Eq. \eqref{Eq:MMa} in the main text.

\subsection{Adaptive QPE with the presence of negative eigenvalues}\label{App:nega}
For a general Hermitian operator $V$ that may contain both positive and negative eigenvalues, the eigenvalues can also be obtained by additional post-processing.
We take the evolution time $\tau_i=2^{m-i}\pi/2\|V\|$, then the positive eigenvalues $v_k>0$ can be estimated with the peak around $\vv_k=v_k/2||V||<1/2$. For the negative eigenvalues, let $\tilde v_j=-v_j/2||V||$ with $0<\tilde a_j<1/2$, then noting that $F_{2^m}(a-\vv_j)=F_{2^m}(a-\vv_j-1)=F_{2^m}[a-(1-\tilde v_j)]$, then the negative eigenvalues $v_j<0$ can be estimated with the peak around $1/2<1-\tilde v_j<1$.
By moving the data of histogram in the range of $1/2<a<1$ to $-1/2<a<0$, and doubling the scale of the $a$ axis, we obtain the possibility distribution of $v_k/||V||$ [see Fig. \ref{Fig:neg}].

\begin{figure}[H]
\centering
  \includegraphics[width=14cm]{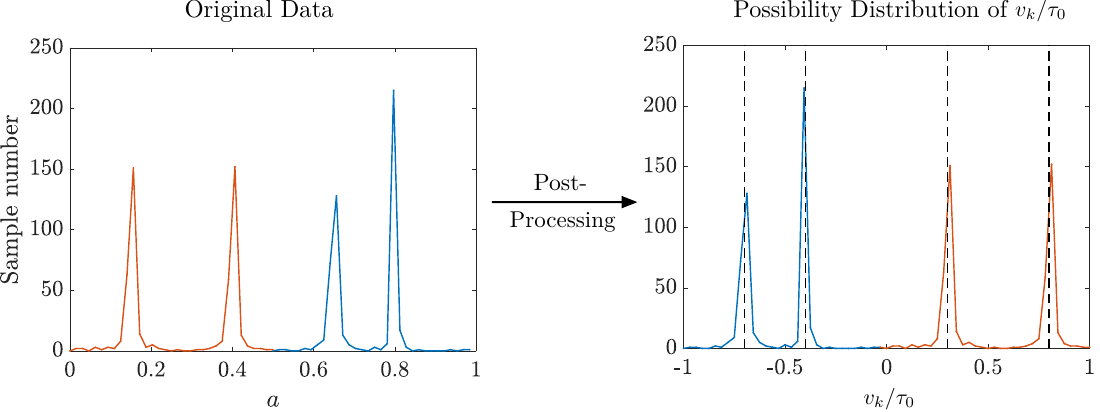}
  \caption{Illustration of adaptive QPE with negative eigenvalues. The possibility distribution of $v_k/\tau_0$ (right panel) is obtained by proper post-processing of original data (left panel), in which $\tau_0=1.25||V||$, then $\tau_i=2^{m-i}\pi/2.5||V||$, see \cite{Note2}. The exact rescaled eigenvalues are indicated by black dashed lines, and the initial state of the target qubits is an equal superposition of all the eigenstates of $V$. The number of RIM rounds is $m=6$ and the simulation contains $10^3$ samples.}\label{Fig:neg}
\end{figure}

\section{QPE for a target system with incoherent noise}\label{incoherent}

\begin{figure}[htbp]
\centering
  \includegraphics[width=16cm]{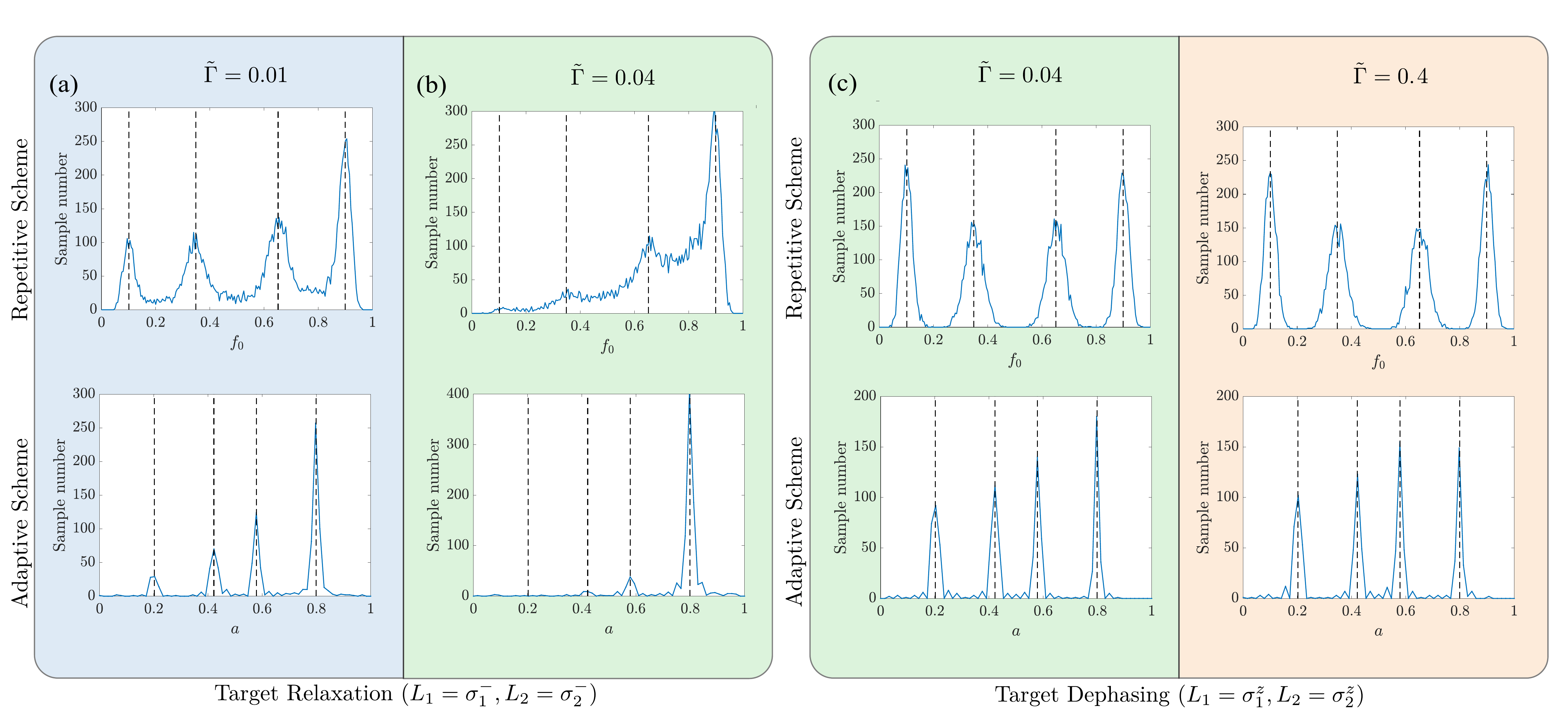}
  \caption{Monte Carlo simulations of QPE of repetitive (upper panel) and adaptive scheme (lower panel) for a spin-star model in the presence of (a-b) relaxation noise and (c-d) dephasing noise with different noise strength. The parameters of relative intensity of noise are (a) $\tilde \Gamma=\Gamma_1/A_1=\Gamma_2/A_2=0.01$, (b-c) $\tilde \Gamma=0.04$, (d) $\tilde \Gamma=0.4$. The black dashed lines show theoretical distribution without noise obtained by spectrum decomposition.  
    We take $10^4$ samples with $m=200$ and $\phi=\pi/2$ for repetitive scheme and $10^3$ samples with $m=6$ for adaptive scheme. }\label{Fig.diss}
\end{figure}

\begin{figure}[htbp]
\centering
  \includegraphics[width=16cm]{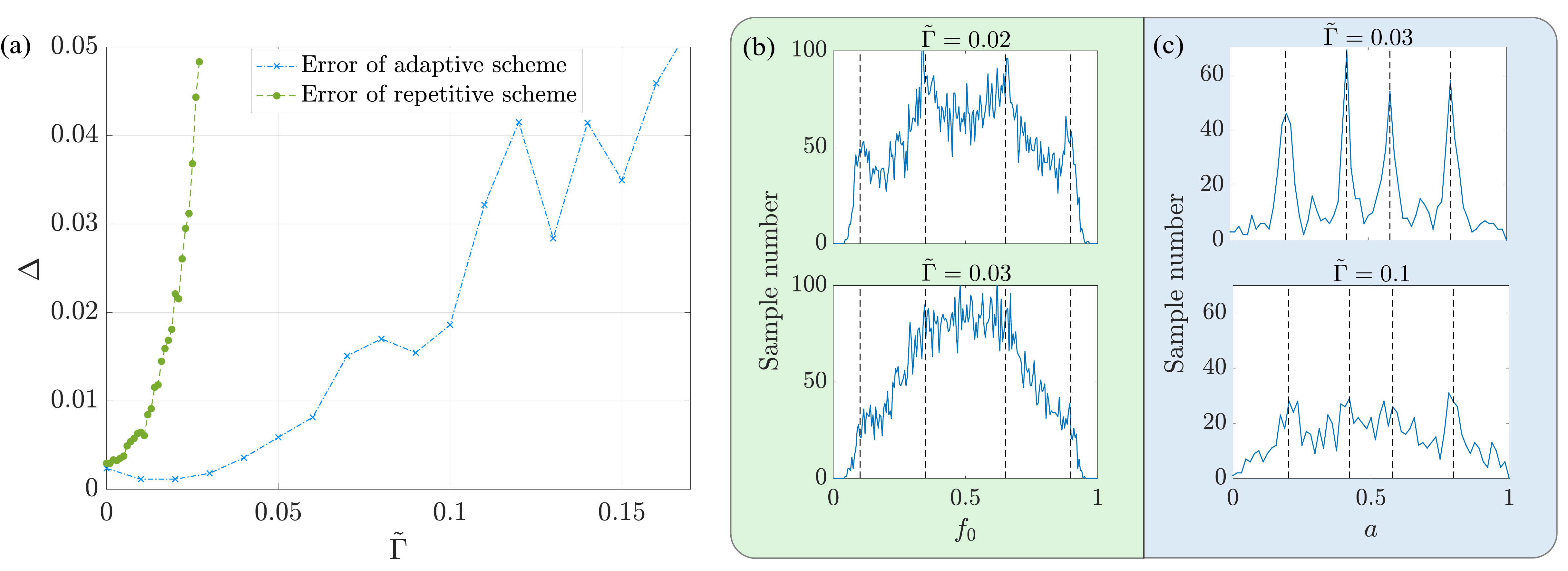}
  \caption{(a)Comparison of noise-resilience and Monte Carlo simulations of (b) repetitive and (c) adaptive QPE schemes for a spin-star model in the presence of relaxation noise ($L_1=\sigma_1^-,L_2=\sigma_2^-,L_3=\sigma_1^+,L_4=\sigma_2^+$).  The black dashed lines in (b-c) show theoretical distribution without noise obtained by spectrum decomposition.
    We take $10^4$ samples with $m=200$ and $\phi=\pi/2$ for repetitive scheme and $10^3$ samples with $m=6$ for adaptive scheme. }\label{Fig.diss2}
\end{figure}

In addition to the coherent noise represented as a perturbation to the Hamiltonian in Eq. \eqref{Hstar}, there may also be incoherent noise on the target system. In this section, we show that both repetitive and adaptive QPE schemes work when the target system suffers dephasing noise and weak relaxation noise.

With incoherent noise on the target system, the evolution of the composite systems can be described by the following Lindblad master equation,
\begin{equation}
  \dv{\rho_{\rm tot}}{t}=-i[H,\rho_{\rm tot}]+\sum_k \Gamma_k \left (L_k\rho_{\rm tot} L_k^{\dagger}-\frac{1}{2}\left\{L_k^{\dagger} L_k,\rho_{\rm tot}\right\}\right),
\end{equation}
where $\rho_{\rm tot}$ is the density matrix of the composite system, $H=\sigma_z\otimes V$ , $L=\sum_{i=1}^K\sigma_i^z$ denotes the target dephasing, $L=\sum_{i=1}^K\sigma_i^-$ with $\sigma^-=|1\rangle\langle0|$ denotes the target relaxation, and $\Gamma_k$ is the dissipation rate.

For Monte Carlo simulations, we use the spin-star model with two target qubits,
\begin{equation}
  V=\sum_{j=1}^2 \frac{A_j}{4}(\bm{\sigma}_{j}\cdot \mathbf{n}_j+\mathbb{I}_j).
\end{equation}
We use $\tilde \Gamma=\Gamma_1/A_1=\Gamma_2/A_2$ to represent the intensity of noise. Then the results show that the dephasing noise of target system does not influence the measurement statistics for both schemes [Fig. \ref{Fig.diss}(c-d)]. However, the relaxation of target system maps $\ket{0}$ to $\ket{1}$, which makes the peaks corresponding to $P_1=\ket{00}\bra{00},\,P_2=\ket{01}\bra{01},\,P_3=\ket{10}\bra{10}$ relax to the peak corresponding to $P_4=\ket{11}\bra{11}$ [Fig. \ref{Fig.diss}(a-b)]. Thus, weak relaxation noise can be tolerated to get a good estimation. 

A single relaxation noise only results in the disappearance of the high-energy peak without altering its position. In other words, the same estimation accuracy can be achieved before the peak vanishes. However, when both types of relaxation noise ($L_1=\sigma_1^-,L_2=\sigma_2^-,L_3=\sigma_2^+,L_4=\sigma_2^+$) are present, the peak position shifts, leading to a decrease in estimation accuracy within noisy systems [see Fig. \ref{Fig.diss2} (b-c)]. In addition, we find that the adaptive scheme exhibits stronger noise resistance compared to the repetitive scheme [see Fig. \ref{Fig.diss2} (a) and compare Fig. \ref{Fig.diss2} (b-c)].

\section{Estimation of the number of samples for iterative QPE}\label{Sec:App.Nsample}
We give a rough estimation of the number of samples needed for both repetitive and adaptive QPE schemes. We get the estimation of eigenvalues from fitting the measurements histogram, which fits the exact probability distribution (see Fig. \ref{MCrep}). Consider any distribution with multiple kernels
\begin{equation}
p(\xi)=\sum_{k=1}^s {\rm Tr}(P_k\rho)K(\xi,v_k),
\end{equation}
where $K$ is the kernel, and $\xi$ is a discrete variable for finite samples ($\xi=m_0/m$ for the repetitive scheme and $\xi=0.a_m\cdots a_2a_1$ for the adaptive scheme). Denote the histogram function from $N$ samples as $h_N(\xi)$, we can use the Hoeffding's inequality to obtain \cite{Hoeffding1963,Roggero2019}
\begin{equation}
  {\rm Pr}(|h_N(\xi)-p(\xi)|\geq \delta)\leq 2e^{-2N\delta^2}=\epsilon.
\end{equation}
This implies that to reach the precision $\delta$ with probability $1-\epsilon$, the sample number is at least
\begin{equation}
  N= \frac{1}{2\delta^2}\ln(\frac{2}{\epsilon}).
\end{equation}
We consider the vicinity of the kernel's peak around $\xi=\xi_k$, where $\xi_k=p_{0k}=[1-\cos(2v_kt+\phi)]/2$ for repetitive QPE and $\xi_k=v_k$ for adaptive QPE. If the difference between the histogram and the exact probability distribution is confined within $\delta$, then the error $\eta$ in estimating $v_k$ can be obtained by solving $K(\xi_k\pm \eta,v_k)=K(\xi_k,v_k)-\delta$.
To compare the two schemes, we expand the Gaussian and Fej\'{e}r kernels at $v_k$. For the Gaussian kernel,
\begin{equation}
  K_G(\xi,v_k) \approx \frac{1}{\sqrt{2\pi\sigma^2}} \left[1 - \frac{(\xi-p_{0k})^2}{2\sigma^2}\right],
\end{equation}
If we take $p_{0k}=1/2$, then $\sigma^2=1/(4m)$, $\delta=\frac{4}{\sqrt{2\pi }}\eta^2m^{3/2}$, then the sample number is
\begin{equation}\label{Ng}
  N_G=\frac{\pi}{16\eta^4 m^3}\ln(\frac{2}{\epsilon})\propto \eta^{-4}t^{-3}\ln(\frac{2}{\epsilon}).
\end{equation}
where the total evolution time is $t=m\tau$ for the repetitive scheme. For the Fej\'{e}r kernel,
\begin{equation}
  K_F(\xi,v_k)\approx 1-\frac{2\pi^2(2^{2m}-1)}{3}(\xi-v_k)^2,
\end{equation}
then $\delta\approx \frac{2\pi^2(2^{2m}-1)}{3}\eta^2$ and the sample number is
\begin{equation}\label{Nf}
  N_F\approx \frac{9}{4\pi^4\eta^42^{4m}}\ln(\frac{2}{\epsilon})\propto \eta^{-4}t^{-4}\ln(\frac{2}{\epsilon}),
\end{equation}
where $t=(2^m-1)\pi$ for the adaptive scheme. Comparing Eq. \eqref{Ng} and Eq. \eqref{Nf}, we find that for the same estimation error $\eta$ and same time consumption $t$, the adaptive scheme needs fewer samples than the repetitive scheme.


\end{widetext}
\newpage

\bibliography{QPE2_zotero}

\end{document}